\documentclass[11pt]{article}
\usepackage{amsmath,amssymb,bbm,amsthm}
\usepackage{fullpage}
\usepackage{thm-restate,color,xcolor,xspace}
\usepackage{hyperref,cleveref}
\usepackage{nicefrac,mathtools}

\newcommand{\f}{\frac}
\newcommand{\cd}{\cdot}

\newcommand{\cds}{\cdots}
\newcommand{\lds}{\ldots}

\newcommand{\sm}{\setminus}
\newcommand{\s}{\subseteq}

\newcommand{\BE}{\begin{enumerate}}
\newcommand{\EE}{\end{enumerate}}
\newcommand{\im}{\item}
\newcommand{\BI}{\begin{itemize}}
\newcommand{\EI}{\end{itemize}}
\def\BAL#1\EAL{\begin{align*}#1\end{align*}}
\def\BALN#1\EALN{\begin{align}#1\end{align}}
\def\BG#1\EG{\begin{gather}#1\end{gather}}

\newcommand{\Prod}{\displaystyle\prod\limits}

\newcommand{\logn}{\log n}

\newcommand{\R}{\mathbb R}

\newcommand{\e}{\epsilon}
\newcommand{\de}{\delta}
\newcommand{\De}{\Delta}
\newcommand{\la}{\lambda}
\newcommand{\g}{\gamma}

\newcommand{\pt}{\partial}
\newcommand{\al}{\alpha}
\newcommand{\be}{\beta}

\newcommand{\Om}{\Omega}

\newcommand{\Th}{\Theta}
\newcommand{\m}{\mathcal}
\newcommand{\ol}{\overline}

\newcommand{\lc}{\lceil}
\newcommand{\rc}{\rceil}

\newcommand{\E}{\mathop{\mathbb E}}

\newcommand{\1}{\mathbbm 1}
\newcommand{\poly}{\textup{poly}}
\newcommand{\polylog}{\textup{polylog}}

\newcommand{\vol}{\textbf{\textup{vol}}}

\newcommand{\lp}{\left(}
\newcommand{\rp}{\right)}
\newcommand{\lb}{\left[}
\newcommand{\rb}{\right]}
\newcommand{\lmt}{\left[\begin{matrix}}
\newcommand{\rmt}{\end{matrix}\right]}

\newtheorem{theorem}{Theorem}[section]
\newtheorem{lemma}[theorem]{Lemma}
\newtheorem{definition}[theorem]{Definition}
\newtheorem{corollary}[theorem]{Corollary}
\newtheorem{observation}[theorem]{Observation}
\newtheorem{claim}[theorem]{Claim}
\newtheorem{subclaim}{Subclaim}
\newtheorem{fact}[theorem]{Fact}
\newtheorem{assumption}[theorem]{Assumption}

\newcommand{\BT}{\begin{theorem}}
\newcommand{\ET}{\end{theorem}}
\newcommand{\BL}{\begin{lemma}}
\newcommand{\EL}{\end{lemma}}
\newcommand{\BD}{\begin{definition}}
\newcommand{\ED}{\end{definition}}
\newcommand{\BC}{\begin{corollary}}
\newcommand{\EC}{\end{corollary}}
\newcommand{\BO}{\begin{observation}}
\newcommand{\EO}{\end{observation}}
\newcommand{\BCL}{\begin{claim}}
\newcommand{\ECL}{\end{claim}}
\newcommand{\BSCL}{\begin{subclaim}}
\newcommand{\ESCL}{\end{subclaim}}
\newcommand{\BF}{\begin{fact}}
\newcommand{\EF}{\end{fact}}
\newcommand{\BA}{\begin{assumption}}
\newcommand{\EA}{\end{assumption}}
\newcommand{\BP}{\begin{proof}}
\newcommand{\EP}{\end{proof}}
\newcommand{\BSP}{\begin{subproof}}
\newcommand{\ESP}{\end{subproof}}
\newcommand{\BPS}{\begin{proof}[Proof (Sketch)]}
\newcommand{\EPS}{\end{proof}}
\Crefname{observation}{Observation}{Observations}
\Crefname{claim}{Claim}{Claims}
\Crefname{subclaim}{Subclaim}{Subclaims}
\Crefname{fact}{Fact}{Facts}
\Crefname{assumption}{Assumption}{Assumptions}

\newenvironment{subproof}[1][\proofname]{%
  \begin{proof}[#1]%
}{%
  \end{proof}%
}

\newcommand{\para}{\paragraph}

\newcommand{\tO}{\widetilde{O}}

\newcommand{\thml}[1]{\label{thm:#1}}
\newcommand{\thm}[1]{\Cref{thm:#1}}
\newcommand{\leml}[1]{\label{lem:#1}}
\newcommand{\lem}[1]{\Cref{lem:#1}}

\newcommand{\clml}[1]{\label{clm:#1}}
\newcommand{\clm}[1]{\Cref{clm:#1}}
\newcommand{\corl}[1]{\label{cor:#1}}
\newcommand{\cor}[1]{\Cref{cor:#1}}

\newcommand{\eqnl}[1]{\label{eq:#1}}
\newcommand{\eqn}[1]{(\ref{eq:#1})}

\newcommand{\secl}[1]{\label{sec:#1}}
\renewcommand{\sec}[1]{\Cref{sec:#1}}

\usepackage[normalem]{ulem}

\begin{document}

\title{Deterministic Mincut in Almost-Linear Time}
\author{Jason Li\thanks{Most of this work was done while the author was an intern at Microsoft Research, Redmond.} \\ Carnegie Mellon University}
\date{\today}
\maketitle

\begin{abstract}
We present a deterministic (global) mincut algorithm for weighted, undirected graphs that runs in $m^{1+o(1)}$ time, answering an open question of Karger from the 1990s. To obtain our result, we de-randomize the construction of the \emph{skeleton} graph in Karger's near-linear time mincut algorithm, which is its only randomized component. In particular, we partially de-randomize the well-known Benczur-Karger graph sparsification technique by random sampling, which we accomplish by the method of pessimistic estimators. Our main technical component is designing an efficient pessimistic estimator to capture the cuts of a graph, which involves harnessing the expander decomposition framework introduced in recent work by Goranci~et~al.~(SODA~2021). As a side-effect, we obtain a structural representation of all approximate mincuts in a graph, which may have future applications.
\end{abstract}

\section{Introduction}

The minimum cut of an undirected, weighted graph 
$G = (V, E, w)$ is a minimum weight subset of edges whose
removal disconnects the graph. Finding the mincut of 
a graph is one of the central problems in combinatorial 
optimization, dating back to the work of 
Gomory and Hu~\cite{gomory1961multi} in 1961 who gave
an algorithm to compute the mincut of an $n$-vertex
graph using $n-1$ max-flow computations. 
Since then, a large body of 
research has been devoted to obtaining faster algorithms 
for this problem. In 1992, Hao and Orlin~\cite{hao1992faster}
gave a clever amortization of the $n-1$ max-flow computations
to match the running time of a single max-flow 
computation. Using the ``push-relabel'' max-flow algorithm 
of Goldberg and Tarjan~\cite{GoldbergT88}, they obtained 
an overall running time of $O(mn \log (n^2/m))$ on 
an $n$-vertex, $m$-edge graph. 
Around the same time, Nagamochi and Ibaraki~\cite{nagamochi1992computing}
(see also \cite{NagamochiI92})
designed an algorithm that bypasses max-flow computations 
altogether, a technique that was further refined by
Stoer and Wagner~\cite{stoer1997simple} (and independently
by Frank in unpublished work).
This alternative method yields a running time of 
$O(mn + n^2 \log n)$. Before 2020, these works yielding a
running time bound of $\tO(mn)$ were the fastest 
{\em deterministic} mincut algorithms for weighted graphs.

Starting with Karger's contraction algorithm in 1993~\cite{karger1993global}, 
a parallel body of work started to emerge in {\em randomized} algorithms
for the mincut problem. This line of work (see also Karger and Stein~\cite{karger1996new})
eventually culminated in a breakthrough paper by Karger~\cite{Karger00}
in 1996 that gave an $O(m\log^3 n)$ time {\em Monte Carlo} algorithm 
for the mincut problem. Note that this algorithm comes to within 
poly-logarithmic factors of the optimal $O(m)$ running time for 
this problem. In that paper, Karger asks whether we can also achieve
near-linear running time using a 
{\em deterministic} algorithm. Even before Karger's work, Gabow~\cite{gabow1995matroid} 
showed that the mincut can be computed in $O(m + \lambda^2 n\log (n^2/m))$
(deterministic) time, where $\lambda$ is the value of the mincut
(assuming integer weights). Note that this result obtains a 
near-linear running time if $\lambda$ is a constant, but 
in general, the running time can be exponential. 
Indeed, for general graphs, Karger's question remains open
after more than 20 years. However, some exciting progress has been
reported in recent years for special cases of this problem. 
In a recent 
breakthrough, Kawarabayashi and Thorup~\cite{KT18}
gave the first near-linear time deterministic algorithm for
this problem {\em for simple graphs}. They obtained a running 
time of $O(m \log^{12} n)$, which was later improved 
by Henzinger, Rao, and Wang~\cite{HenzingerRW17} to 
$O(m \log^2 n\log\log^2 n)$, and then simplified by
Saranurak~\cite{saranurak2021simple} at the cost of $m^{1+o(1)}$ running time.
From a technical perspective, 
Kawarabayashi and Thorup's work introduced the idea of using low conductance 
cuts to find the mincut of the graph, a very powerful idea that 
we also exploit in this paper.

In 2020, the author, together with Debmalya Panigrahi~\cite{LP20}, obtained the first improvement to deterministic mincut for weighted graphs since the 1990s, obtaining a running time of $O(m^{1+\e})$ plus polylogarithmic calls to a deterministic exact $s$--$t$ max-flow algorithm. Using the fastest deterministic algorithm for weighted graphs of Goldberg~and~Rao~\cite{goldberg1998beyond}, their running time becomes $\tO(m^{1.5})$.\footnote{In this paper, $\tO(\cd)$ notation hides polylogarithmic factors in $n$, the number of vertices of the graph.} Their algorithm was inspired by the conductance-based ideas of Kawarabayashi and Thorup  and introduced expander decompositions into the scene. While it is believed that a near-linear time algorithm exists for $s$--$t$ max-flow---which, if deterministic, would imply a near-linear time algorithm for deterministic mincut---the best max-flow algorithms, even for unweighted graphs, is still $m^{4/3+o(1)}$~\cite{liu2020faster}. For the deterministic, weighted case, no improvement since Goldberg-Rao~\cite{goldberg1998beyond} is known.

The main result of this paper is a new deterministic algorithm for mincut that does not rely on $s$--$t$ max-flow computations and achieves a running time of $m^{1+o(1)}$, answering Karger's open question. 

\BT\thml{main}
There is a deterministic mincut algorithm for weighted, undirected graphs that runs in $m^{1+o(1)}$ time.
\ET

\subsection{Our Techniques}
Our approach differs fundamentally from the one in~\cite{LP20} that relies on $s$--$t$ max-flow computations.
At a high level, we follow Karger's approach and essentially de-randomize the single randomized procedure in Karger's near-linear time mincut algorithm~\cite{Karger00}, namely the construction of the \emph{skeleton} graph, which Karger accomplishes through the Benczur-Karger graph sparsification technique by random sampling. We remark that
our de-randomization does not recover a full $(1+\e)$-approximate graph sparsifier, but the skeleton graph that we obtain is sufficient to solve the mincut problem.

Let us first briefly review the Benczur-Karger graph sparsification technique, and discuss the difficulties one encounters when trying to de-randomize it. Given a weighted, undirected graph, the sparsification algorithm samples each edge independently with a probability depending on the weight of the edge and the global mincut of the graph, and then re-weights the sampled edge accordingly. In traditional graph sparsification, we require that every cut in the graph has its weight preserved up to a $(1+\e)$ factor. There are exponentially many cuts in a graph, so a naive union bound over all cuts does not work. Benczur and Karger's main insight is to set up a more refined union bound, layering the (exponentially many) cuts in a graph by their weight. They show that for all $\al\ge1$, there are only $n^{c\al}$ many cuts in a graph whose weight is roughly $\al$ times the mincut, and each one is preserved up to a $(1+\e)$ factor with probability $1-n^{-c'\al}$, for some constants $c'\gg c$. In other words, they establish a union bound layered by the $\al$-approximate mincuts of a graph, for each $\al\ge1$.

One popular method to de-randomize random sampling algorithms is through \emph{pessimistic estimators}, which is a generalization of the well-known method of conditional probabilities. For the graph sparsification problem, the method of pessimistic estimators can be implemented as follows. The algorithm considers each edge one by one in some arbitrary order, and decides on the spot whether to keep or discard each edge for the sparsifier. To make this decision, the algorithm maintains a \emph{pessimistic estimator}, which is a real number in the range $[0,1)$ that represents an upper bound on the probability of failure should the remaining undecided edges each be sampled independently at random. In many cases, the pessimistic estimator is exactly the probability upper bound that one derives from analyzing the random sampling algorithm, except conditioned on the edges kept and discarded so far. The algorithm makes the choice---whether to keep or discard the current edge---based on whichever outcome does not increase the pessimistic estimator; such a choice must always exist for the pessimistic estimator to be valid. Once all edges are processed, the pessimistic estimator must still be a real number less than $1$. But now, since there are no more undecided edges, the probability of failure is either $0$ or $1$. Since the pessimistic estimator is an upper bound which is less than $1$, the probability of failure must be $0$; in other words, the set of chosen edges is indeed a sparsifier of the graph.

In order for this de-randomization procedure to be efficient, the pessimistic estimator must be quickly evaluated and updated after considering each edge. Unfortunately, the probability union bound in the Benczur-Karger analysis involves all cuts in the graph, and is therefore an expression of exponential size and too expensive to serve as our pessimistic estimator. To design a more efficient pessimistic estimator, we need a more compact, easy-to-compute union bound over all cuts of the graph. We accomplish this by grouping all cuts of the graph into two types: small cuts and large cuts.

\para{Small cuts.}
Recall that our goal is to preserve cuts in the graph up to a $(1+\e)$ factor.
Let us first restrict ourselves to all $\al$-approximate mincuts of the graph for some $\al=n^{o(1)}$. There can be $n^{\Om(\al)}$ many such cuts, so the naive union bound is still too slow. Here, our main strategy is to establish a \emph{structural representation} of all $\al$-approximate mincuts of a graph, with the goal of deriving a more compact ``union bound" over all $\al$-approximate cuts. This structure is built from an \emph{expander hierarchy} of the graph, which is a hierarchical partitioning of the graph into disjoint expanders introduced by Goranci~et~al.~\cite{goranci2020expander}. The connection between expanders and the mincut problem has been observed before~\cite{KT18,LP20}: in an expander with conductance $\phi$, all $\al$-approximate mincuts must have at most $\al/\phi$ vertices on one side, so a compact representation is simply all cuts with at most $\al/\phi$ vertices on one side. Motivated by this connection, we show that if the original graph is itself an expander, then it is enough to preserve all vertex degrees and all edge weights up to an additive $\e'\la$ factor, where $\la$ is the mincut of the graph and $\e'$ depends on $\e,\al,\phi$. 
We present the unweighted expander case in \sec{expander} as a warm-up, which features all of our ideas except for the final expander decomposition step. To handle general graphs, we exploit the full machinery of the expander hierarchy~\cite{goranci2020expander}.

\para{Large cuts.}
For the large cuts---those that are not $\al$-approximate mincuts---our strategy differs from the pessimistic estimator approach. Here, our aim is not to preserve each of them up to a $(1+\e)$-factor, but a $\g$-factor for a different parameter $\g=n^{o(1)}$. This relaxation prevents us from obtaining a full $(1+\e)$-approximate sparsification of the graph, but it still works for the mincut problem as long as the large cuts do not fall below the original mincut value. While a deterministic $(1+\e)$-approximate sparsification algorithm in near-linear time is unknown, one exists for $\g$-approximation sparsification for some $\g=n^{o(1)}$~\cite{CGL19}. In our case, we actually need the sparsifier to be \emph{uniformly weighted}, so we construct our own sparsifier in \sec{lossy}, again via the expander hierarchy. Note that if the original graph is an expander, then we can take any expander whose degrees are roughly the same; in particular, the sparsifier does not need to be a subgraph of the original graph. To summarize, for the large cuts case, we simply construct an $\g$-approximate sparsifier deterministically, bypassing the need to de-randomize the Benczur-Karger random sampling technique.

\para{Combining them together.}
Of course, this $\g$-approximate sparsifier destroys the guarantee of the small cuts, which need to be preserved $(1+\e)$-approximately. Our strategy is to combine the small cut sparsifier and the large cut sparsifier together in the following way. We take the union of the small cut sparsifier with a ``lightly'' weighted version of the large cut sparsifier, where each edge in it is weighted by $\e/\g$ times its normal weight. This way, each small cut of weight $w$ suffers at most an additive $\g w\cd \e/\g = \e w$ weight from the ``light'' large cut sparsifier, so we do not destroy the small cuts guarantee (up to replacing $\e$ with $2\e$). Moreover, each large cut of weight $w\ge\al\la$ is weighted by at least $w/\g \cd \e/\g \ge \al\la/\g \cd \e/\g = \al/\g^2 \cd \e\la$, where $\la$ is the mincut of the original graph. Hence, as long as $\al\ge\g^2/\e$, the large cuts have weight at least the mincut, and the property for large cuts is preserved.

\para{Unbalanced vs.\ balanced.}
We remark that our actual separation between small cuts and large cuts is somewhat different; we use \emph{unbalanced} and \emph{balanced} instead to emphasize this distinction. Nevertheless, we should intuitively think of unbalanced cuts as having small weight and balanced as having large weight; rather, the line is not drawn precisely at a weight threshold of $\al\la$. The actual separation is more technical, so we omit it in this overview section.

\subsection{Preliminaries}

In this paper, all graphs are undirected, and $n$ and $m$ denote the number of vertices and edges of the input graph in question. All graphs are either unweighted or weighted multigraphs with polynomially bounded edge weights, i.e., in the range $[\f1{\poly(n)},\poly(n)]$. \textbf{We emphasize that even \emph{weighted} graphs are multigraphs, which we find more convenient to work with.} 

We begin with more standard notation. For an unweighted graph $G=(V,E)$ and vertices $u,v\in V$, let $\#(u,v)$ be the number of edges $e\in E$ with endpoints $u$ and $v$. For a weighted graph $G=(V,E)$ and edge $e\in E$, let $w(e)$ be the weight of the edge, and for vertices $u,v\in V$, let $w(u,v)$ be the \emph{sum} of the weights $w(e)$ of all (parallel) edges $e$ between $u$ and $v$. For disjoint sets of vertices $S,T\s V$, define $E(S,T)\s E$ as the set of edges with one endpoint in $S$ and the other in $T$, and define $\pt S:=E(S,V\sm S)$. For a set $F\s E$ of edges, denote its cardinality by $|F|$ if $G$ is unweighted, and its total weight by $w(F)$ if $G$ is weighted. Define the \emph{degree} $\deg(v)$ of vertex $v\in V$ to be $|\pt(\{v\})|$ if $G$ is unweighted, and $w(\pt(\{v\}))$ if $G$ is weighted. For a set $S\s V$, define $\vol(S):=\sum_{v\in S}\deg(v)$. A \emph{cut} of $G$ is the set of edges $\pt S$ for some $\emptyset\subsetneq S\subsetneq V$, and the \emph{mincut} of $G$ is the cut $\pt S$ in $G$ that minimizes $|\pt S|$ or $w(\pt S)$ depending on if $G$ is unweighted or weighted. When the graph $G$ is ambiguous, we may add a subscript of $G$ in our notation, such as $\#_G(u,v)$.

\subsubsection{Karger's Approach}

In this section, we outline Karger's approach to his near-linear time randomized mincut algorithm and set up the necessary theorems for our deterministic result. Karger's algorithm has two main steps. First, it computes a small set of (unweighted) trees on vertex set $V$ such that the mincut \emph{$2$-respects} one of the trees $T$, defined as follows:
\BD
Given a weighted graph $G$ and an unweighted tree $T$ on the same set of vertices, a cut $\pt_G S$ \emph{$2$-respects} the tree $T$ if $|\pt_TS|\le2$.
\ED
Karger accomplishes this goal by first \emph{sparsifying} the graph into an unweighted \emph{skeleton} graph using the well-known Benzcur-Karger sparsification by random sampling, and then running a \emph{tree packing} algorithm of Gabow~\cite{gabow1995matroid} on the skeleton graph.
\BT[Karger~\cite{Karger00}]\thml{skeleton}
Let $G$ be a weighted graph, let $m'$ and $c'$ be parameters, and let $H$ be an unweighted graph on the same vertices, called the \emph{skeleton} graph, with the following properties:
 \BE
 \im[\textup{(a)}] $H$ has $m'$ edges,
 \im[\textup{(b)}] The mincut of $H$ is $c'$, and
 \im[\textup{(c)}] The mincut in $G$ corresponds (under the same vertex partition) to a $\nicefrac76$-approximate mincut in $H$.
 \EE
Given graphs $G$ and $H$, there is a deterministic algorithm in $O(c'm'\logn)$ time that constructs $O(c')$ trees on the same vertices such that one of them $2$-respects the mincut in $G$.
\ET

The second main step of Karger's algorithm is to compute the mincut of $G$ given a tree that $2$-respects the mincut. This step is deterministic and is based on dynamic programming.
\BT[Karger~\cite{Karger00}]\thml{respect}
Given a weighted, undirected graph $G$ and a (not necessarily spanning) tree $T$ on the same vertices, there is a deterministic algorithm in $O(m\log^2n)$ time that computes the minimum-weight cut in $G$ that $2$-respects the tree $T$.
\ET

Our main technical contribution is a deterministic construction of the skeleton graph used in \thm{skeleton}.
Instead of designing an algorithm to produce the skeleton graph directly, it is more convenient to prove the following, which implies a skeleton graph by the following claim.
\begin{restatable}{theorem}{Sparsifier}\thml{sparsifier}
For any $0<\e\le1$, we can compute, in deterministic $\e^{-7}2^{O(\logn)^{5/6}(\log\logn)^{O(1)}}m$ time, an unweighted graph $H$ and some weight $W=\e^7\la/2^{O(\logn)^{5/6}(\log\logn)^{O(1)}}$ such that
 \BE
 \im For any mincut $\pt S^*$ of $G$, we have $W\cd|\pt_HS^*|\le(1+\e)\la$, and
 \im For any cut $\emptyset\subsetneq S\subsetneq V$ of $G$, we have $W\cd|\pt_HS| \ge (1-\e)\la$.
 \EE
\end{restatable}

\BCL\clml{satisfy}
For $\e=0.01$, the graph $H$ in \thm{sparsifier} fulfills the conditions of \thm{skeleton} with $m'=m^{1+o(1)}$ and $c'=n^{o(1)}$.
\ECL
\BP
Since the algorithm of \thm{sparsifier} takes $m^{1+o(1)}$ time, the output graph $H$ must have $m^{1+o(1)}$ edges, fulfilling condition~(a) of \thm{skeleton}. For any mincut $S^*$ of $G$, by property~(1) of \thm{sparsifier},
we have $|\pt_HS^*|\le (1+\e)\la/W\le n^{o(1)}$, fulfilling condition~(b).  
For any cut $\emptyset\subsetneq S\subsetneq V$, by property~(2), we have $|\pt_HS|\ge(1-\e)\la/W$. In other words, $S^*$ is a $(1+\e)/(1-\e)$-approximate mincut, which is a $\nicefrac76$-approximate mincut for $\e=0.01$, fulfilling condition~(c). 
\EP

With the above three statements in hand, we now prove \thm{main} following Karger's approach. Run the algorithm of \thm{sparsifier} to produce a graph $H$ which, by \clm{satisfy}, satisfies the conditions of \thm{skeleton}. Apply \thm{skeleton} on $G$ and the skeleton graph $H$, producing $n^{o(1)}$ many trees such that one of them $2$-respects the mincut in $G$. Finally, run \thm{respect} on each tree separately and output the minimum $2$-respecting cut found among all the trees, which must be the mincut in $G$. Each step requires $2^{O(\logn)^{5/6}(\log\logn)^{O(1)}}m$ deterministic time, proving \thm{main}.

Thus, the main focus for the rest of the paper is proving \thm{sparsifier}.

\subsubsection{Spectral Graph Theory}

Central to our approach are the well-known concepts of \emph{conductance}, \emph{expanders}, and the graph \emph{Laplacian} from spectral graph theory. 

\BD[Conductance, expander]
The \emph{conductance} of a weighted graph $G$ is 
\[ \Phi(G):= \min_{\emptyset\subsetneq S\subsetneq V} \f{w(E(S, V\sm S))}{\min\{\vol(S),\vol(V\sm S)\}} .\]
For the conductance of an unweighted graph, replace $w(E(S,V\sm S))$ by $|E(S,V\sm S)|$. We say that $G$ is a \emph{$\phi$-expander} if $\Phi(G)\ge\phi$.
\ED

\BD[Laplacian]
The \emph{Laplacian} $L_G$ of a weighted graph $G=(V,E)$ is the $n\times n$ matrix, indexed by $V\times V$, where
 \BE
 \im[\textup{(a)}] Each diagonal entry $(v,v)$ has entry $\deg(v)$, and
 \im[\textup{(b)}] Each off-diagonal entry $(u,v)$ ($u\ne v$) has weight $-w(u,v)$ if $(u,v)\in E$ and $0$ otherwise.
 \EE
\ED

The only fact we will use about Laplacians is the following well-known fact, that cuts in graphs have the following nice form:
\BF\label{fact:L}
For any weighted graph $G=(V,E)$ with Laplacian $L_G$, and for any subset $S\s V$, we have 
\[ w(\pt S) = \1_S^TL_G\1_S ,\]
where $\1_S\in\{0,1\}^V$ is the vector with value $1$ at vertex $v$ if $v\in S$, and value $0$ otherwise. For unweighted graph $G$, replace $w(\pt S)$ with $|\pt S|$.
\EF


\section{Expander Case}\secl{expander}

In this section, we prove \thm{sparsifier} restricted to the case when $G$ is an \emph{unweighted expander}. Our aim is to present an informal, intuitive exposition that highlights our main ideas in a relatively simple setting. Since this section is not technically required for the main result, we do not attempt to formalize our arguments, deferring the rigorous proofs to the general case in \sec{general}.

\BT\thml{exp-case}
Let $G$ be an unweighted $\phi$-expander multigraph. For any $0<\e\le1$, we can compute, in deterministic $m^{1+o(1)}$ time, an unweighted graph $H$ and some weight $W=\e^3\la/n^{o(1)}$ such that
 \BE
 \im[\textup{(a)}] For any mincut $\pt_G S^*$ of $G$, we have $W\cd|\pt_HS^*|\le(1+\e)\la$, and
 \im[\textup{(b)}] For any cut $\pt_G S$ of $G$, we have $W\cd|\pt_HS| \ge (1-\e)\la$.
 \EE
\ET

For the rest of this section, we prove \thm{exp-case}. 

Consider an arbitrary cut $\pt_G S$. By \Cref{fact:L}, we have
\BG  |\pt_G S| = \1_S^TL_G\1_S = \lp\sum_{v\in S}\1_v^T\rp L_G\lp\sum_{v\in S}\1_v\rp = \sum_{u,v\in S}\1_u^TL_G\1_v . \eqnl{sumS}\EG
Suppose we can approximate each $\1_u^TL_G\1_v$ to an additive error of $\e'\la$ for some small $\e'$ (depending on $\e$); that is, suppose that our graph $H$ and weight $W$ satisfy 
\[ |\1^T_uL_G\1_v - W\cd\1_u^TL_H\1_v|\le\e'\la  \]
for all $u,v\in V$.
Then, by \eqn{sumS}, we can approximate $|\pt_G S|$ up to an additive $|S|^2\e'\la$, or a multiplicative $(1+|S|^2\e')$, which is good if $|S|$ is small. Similarly, if $|V\sm S|$ is small, then we can replace $S$ with $V\sm S$ in \eqn{sumS} and approximate $|\pt_GS|=|\pt_G(V\sm S)|$ to the same factor. Motivated by this observation, we define a set $S\s V$ to be \emph{unbalanced} if $\min\{ \vol(S),\vol(V\sm S)\}\le\al\la/\phi$ for some $\al=n^{o(1)}$ to be set later. Similarly, define a cut $\pt_G S$ to be unbalanced if the set $S$ is unbalanced.
Note that an unbalanced set $S$ must have either $|S|\le\al/\phi$ or $|V\sm S|\le\al/\phi$, since if we assume without loss of generality that $\vol(S)\le\vol(V\sm S)$, then
\BG |S|\la\le\sum_{v\in S}\deg(v)=\vol(S)\le\al\la/\phi ,\eqnl{unbal}\EG
where the first inequality uses that each degree cut $\pt(\{v\})$ has weight $\deg(v)\ge\la$. Moreover, since $G$ is a $\phi$-expander, the mincut $\pt_GS^*$ is unbalanced because, assuming without loss of generality that $\vol(S^*)\le\vol(V\sm S^*)$, we obtain
\[ \f{|\pt_G(S^*)|}{\vol(S^*)}\ge\Phi(G)\ge\phi\implies \vol(S^*)\le\la/\phi\le\al\la/\phi .\]

To approximate all unbalanced cuts, it suffices by \eqn{sumS}~and~\eqn{unbal} to approximate each $\1^T_uL_G\1_v$ up to additive error $(\phi/\al)^2\e\la$. 
When $u\ne v$, the expression $\1_u^TL_G\1_v$ is simply the negative of the number of parallel $(u,v)$ edges in $G$. So, approximating $\1_u^TL_G\1_v$ up to additive error $\e\la$ simply amounts to approximating the number of parallel $(u,v)$ edges. When $u=v$, the expression $\1_v^TL_G\1_v$ is simply the degree of $v$, so approximating it amounts to approximating the degree of $v$. 

Consider what happens if we randomly sample each edge with probability $p=\Th(\f{\al\logn}{\e^2\phi\la})$ and weight the sampled edges by $\widehat W:=1/p$ to form the sampled graph $\widehat H$. For the terms $\1_u^TL_G\1_v$ ($u\ne v$), we have $\#_G(u,v)\le\vol(S)\le\al\la/\phi$. Let us assume for simplicity that $\#_G(u,v)=\al\la/\phi$, which turns out to be the worst case. By Chernoff bounds, for $\de=\e\phi/\al$,
\BALN
 \Pr\lb \left| \#_{\widehat H}(u,v) - p\cd \#_G(u,v) \right| > \de\cd p\cd \#_G(u,v)\rb &< 2\exp(-\de^2 \cd p \cd \#_G(u,v)/3) \nonumber
\\&= 2\exp\lp -\lp\f{\e\phi}\al\rp^2\cd \Th\left(\f{\al\logn}{\e^2\phi\la}\right) \cd \f{\al\la/\phi}3 \rp \eqnl{pessimistic}
\\&=2\exp(-\Th(\logn)) ,\nonumber
\EALN
which we can set to be much less than $1/n^2$. We then have the implication
\[ \left|\#_{\widehat H}(u,v) - p\cd \#_G(u,v)\right| \le \de\cd p\cd \#_G(u,v) \implies \left|\1_u^T(L_G-L_{\widehat H})\1_v\right| \le \de\cd \#_G(u,v)=\e\phi/\al \cd \al\la/\phi = \e\la .\]
 Similarly, for the terms $\1_v^TL_G\1_v$, we have $\deg(v)\le\vol(S)\le\al\la/\phi$, and the same calculation can be made.

From this random sampling analysis, we can derive the following pessimistic estimator. Initially, it is the sum of the quantities \eqn{pessimistic} for all $(u,v)$ satisfying either $u=v$ or $(u,v)\in E$. This sum has $O(m)$ terms which sum to less than $1$, so it can be efficiently computed and satisfies the initial condition of a pessimistic estimator. After some edges have been considered, the probability upper bounds \eqn{pessimistic} are modified to be conditional to the choices of edges so far, which can still be efficiently computed. At the end, for each unbalanced set $S$, the graph $\widehat H$ will satisfy
\[ \big| |\pt_G S| - \widehat W\cd|\pt_{\widehat H}S| \big| \le \e\la \implies (1-\e) |\pt_GS| \le \widehat W\cd|\pt_{\widehat H}S| \le (1+\e)|\pt_GS| .\]
Since any mincut $\pt_GS^*$ is unbalanced, we fulfill condition~(a) of \thm{exp-case}. We also fulfill condition~(b) for any cut with a side that is unbalanced. This concludes the unbalanced case; we omit the rest of the details, deferring the pessimistic estimator and its efficient computation to the general case, specifically \sec{random}.

Define a cut to be \emph{balanced} if it is not unbalanced. For the balanced cuts, it remains to fulfill condition~(b), which may not hold for the graph $\widehat H$. Our solution is to ``overlay'' a fixed expander onto the graph $\widehat H$, weighted small enough to barely affect the mincut (in order to preserve condition~(a)), but large enough to force all balanced cuts to have weight at least $\la$. In particular, let $\widetilde H$ be an unweighted $\Th(1)$-expander on the same vertex set $V$ where each vertex $v\in V$ has degree $\Th(\deg_G(v)/\la)$, and let $\widetilde W:=\Th(\e\phi\la)$. We should think of $\widetilde H$ as a ``lossy" sparsifier of $G$, in that it approximates cuts up to factor $O(1/\phi)$, not $(1+\e)$.

Consider taking the ``union'' of the graph $\widehat H$ weighted by $\widehat W$ and the graph $\widetilde H$ weighted by $\widetilde W$. More formally, consider a weighted graph $H'$ where each edge $(u,v)$ is weighted by $\widehat W\cd w_{\widehat H}(u,v)+\widetilde W\cd w_{\widetilde H}(u,v)$. We now show two properties: (1) the mincut gains relatively little weight from $\widetilde H$ in the union $H'$, and (2) any balanced cut automatically has at least $\la$ total weight from $\widetilde H$.
\BE
\im For a mincut $\pt _GS^*$ in $G$ with $\vol_G(S^*)\le |\pt _G S^*|/\phi=\la/\phi$, the cut crosses
\[  w(\pt_{\widehat H}S^*)\le\vol_{\widehat H}(S^*)\le \Th(1)\cd\vol_G(S^*)/\la\le \Th(1/\phi) \]
 edges in $\widetilde H$, for a total cost of at most $\Th(1/\phi)\cd\Th(\e\phi\la) \le \e\la$. 
\im For a balanced cut $\pt _GS$, it satisfies $ |\pt_GS|\ge\phi\cd\vol_G(S)\ge\al\la$, so it crosses 
\[  w(\pt_{\widehat H}S)\ge \Th(1)\cd\vol_{\widehat H}(S) \ge \Th(1)\cd\vol_G(S)/\la \ge \Th(\al/\phi) \]
 many edges in $\widetilde H$, for a total cost of at least $\Th(\al/\phi) \cd \Th(\e\phi\la)$. Setting $\al := \Th(\f1\e)$, the cost becomes at least $\la$.
\EE
Therefore, in the weighted graph $H'$, the mincut has weight at most $(1+O(\e))\la$, and any cut has weight at least $(1-\e)\la$. We can reset $\e$ to be a constant factor smaller so that the factor $(1+O(\e))$ becomes $(1+\e)$.

To finish the proof of \thm{exp-case}, it remains to extract an unweighted graph $H$ and a weight $W$ from the weighted graph $H'$. Since $\widehat W=\Th(\f{\e^2\phi\la}{\al\logn})=\Th(\f{\e^3\phi\la}\logn)$ and $\widetilde W=\Th(\e\phi\la)$, we can make $\widetilde W$ an integer multiple of $\widehat W$, so that each edge in $H'$ is an integer multiple of $\widehat W$. We can therefore set $W:=\widehat W$ and define the unweighted graph $H$ so that $\#_H(u,v)=w_{H'}(u,v)/\widehat W$ for all $u,v\in V$.

\section{General Case}\secl{general}

This section is dedicated to proving \thm{sparsifier}. For simplicity, we instead prove the following restricted version first, which has the additional assumption that the maximum edge weight in $G$ is bounded. At the end of this section, we show why this assumption can be removed to obtain the full \thm{sparsifier}.

\begin{restatable}{theorem}{SpRestricted}\thml{sparsifier-restricted}
There exists a function $f(n)\le 2^{O(\logn)^{5/6}(\log\logn)^{O(1)}}$ such that the following holds. Let $G$ be a graph with mincut $\la$ \textbf{and maximum edge weight at most $\e^7\la/f(n)$.}
For any $0<\e\le1$, we can compute, in deterministic $2^{O(\logn)^{5/6}(\log\logn)^{O(1)}}m$ time, an unweighted graph $H$ and some weight $W\ge\e^7\la/f(n)$ such that the two properties of \thm{sparsifier} hold, i.e.,
 \BE
 \im For any mincut $S^*$ of $G$, we have $W\cd|\pt_HS^*|\le(1+\e)\la$, and
 \im For any cut $\emptyset\subsetneq S\subsetneq V$ of $G$, we have $W\cd|\pt_HS| \ge (1-\e)\la$.
 \EE
\end{restatable}

\subsection{Expander Decomposition Preliminaries}\secl{exp-prelim}

Our main tool in generalizing the expander case is \emph{expander decompositions}, which was popularized by Spielman~and~Teng~\cite{SpielmanT04} and is quickly gaining traction in the area of fast graph algorithms. The general approach to utilizing expander decompositions is as follows. First, solve the case when the input graph is an expander, which we have done in \sec{expander} for the problem described in \thm{sparsifier}. Then, for a general graph, \emph{decompose} it into a collection of expanders with few edges between the expanders, solve the problem each expander separately, and combine the solutions together, which often involves a recursive call on a graph that is a constant-factor smaller. For our purposes, we use a slightly stronger variant than the usual expander decomposition that ensures \emph{boundary-linkedness}, which will be important in our analysis. The following definition is inspired by~\cite{goranci2020expander}; note that our variant is weaker than the one in Definition~4.2 of~\cite{goranci2020expander} in that we only guarantee their property~(2). For completeness, we include a full proof in \sec{exp-decomp} that is similar to the one in~\cite{goranci2020expander}, and assuming a subroutine called \textsf{WeightedBalCutPrune} from~\cite{LS21}.


\BT[Boundary-linked expander decomposition]\thml{exp-decomp}
Let $G=(V,E)$ be a graph and let $r\ge1$ be a parameter. There is a deterministic algorithm in $m^{1+O(1/r)} + \tO(m/\phi^2)$ time that, for any parameters $\be \le (\logn)^{-O(r^4)}$ and $\phi\le\be$, partitions $V=V_1\uplus\cds\uplus V_k$ such that
 \BE
 \im Each vertex set $V_i$ satisfies
\BG \min_{\emptyset\subsetneq S\subsetneq V_i} \f{ w(\pt_{G[V_i]}S)}{\min\{\vol_{G[V_i]}(S) + \f\be\phi w(E_G(S,V\sm V_i)),\vol_{G[V_i]}(V_i\sm S) + \f\be\phi w(E_G(V_i\sm S,V\sm V_i))\}} \ge \phi . \eqnl{exp} \EG
Informally, we call the graph $G[V_i]$ together with its boundary edges $E_G(V_i,V\sm V_i)$ a $\be$-\emph{boundary-linked} $\phi$-expander.\footnote{For unweighted graphs, \cite{goranci2020expander} uses the notation $G[V_i]^{\be/\phi}$ to represent a graph where each (boundary) edge in $E(V_i,V\sm V_i)$ is replaced with $\be/\phi$ many self-loops at the endpoint in $V_i$. With this definition, \eqn{exp} is equivalent to saying that $G[V_i]^{\be/\phi}$ is a $\phi$-expander. We will use this definition when proving \thm{exp-decomp} in \sec{exp-decomp}.}
In particular, for any $S$ satisfying 
\[ \vol_{G[V_i]}(S) + \f\be\phi w(E_G(S,V\sm V_i)) \le \vol_{G[V_i]}(V_i\sm S) + \f\be\phi w(E_G(V_i\sm S,V\sm V_i)),\]
 we simultaneously obtain 
\[ \f{ w(\pt_{G[V_i]}S)}{\vol_{G[V_i]}(S)} \ge \phi \qquad\text{and}\qquad \f{ w(\pt_{G[V_i]}S)}{ \f\be\phi w(E_G(S,V\sm V_i))} \ge \phi \iff  \f{ w(\pt_{G[V_i]}S)}{  w(E_G(S,V\sm V_i))} \ge \be.\]
The right-most inequality is where the name ``boundary-linked" comes from.
 \im The total weight of ``inter-cluster" edges, $ w(\pt V_1\cup\cds\cup \pt V_k)$, is at most $ (\logn)^{O(r^4)}\phi\vol(V)$.
 \EE
\ET

Note that for our applications, it's important that the boundary-linked parameter $\be$ is much larger than $\phi$. This is because in our recursive algorithm, the approximation factor will blow up by roughly $1/\be$ per recursion level, while the instance size shrinks by roughly $\phi$.

In order to capture recursion via expander decompositions, we now define a \emph{boundary-linked expander decomposition sequence} $\{G^i\}$ on the graph $G$ in a similar way to~\cite{goranci2020expander}. Compute a boundary-linked expander decomposition for $\be$ and $\phi\le\be$ to be determined later, contract each expander,\footnote{Since we are working with weighted multigraphs, we do \emph{not} collapse parallel edges obtained from contraction into single edges.} and recursively decompose the contracted graph until the graph consists of a single vertex. Let $G^0=G$ be the original graph and $G^1,G^2,\lds,G^L$ be the recursive contracted graphs. Note that each graph $G^i$ has minimum degree at least $\la$, since any degree cut in any $G^i$ induces a cut in the original graph $G$. Each time we contract, we will keep edge identities for the edges that survive, so that $E(G^0) \supseteq E(G^1) \supseteq \cds\supseteq E(G^L)$. Let $U^i$ be the vertices of $G^i$.

For the rest of \sec{exp-prelim}, fix an expander decomposition sequence $\{G^i\}$ of $G$.
For any subset $\emptyset\subsetneq S\subsetneq V$, we now define an \emph{decomposition sequence} of $S$ as follows. Let $S^0=S$, and for each $i>0$, construct $S^{i+1}$ as a subset of the vertices of $G^{i+1}$, as follows. Take the expander decomposition of $G^i$, which partitions the vertices $U^i$ of $G^i$ into, say, $U^i_1,\lds, U^i_{k_i}$. Each of the $U^i_j$ gets contracted to a single vertex $u_j$ in $G^{i+1}$. For each $U^i_j$, we have a choice whether to add $u_j$ to $S^{i+1}$ or not. This completes the construction of $S^{i+1}$.  Define the ``difference'' $D^i_j=U^i_j\sm S^i$ if $u_j\in S^i$, and $D^i_j=U^i_j\cap S^i$ otherwise.  The sets $S^i$, $U^i_j$, and $D^i_j$ define the decomposition sequence of $S$.

We now prove some key properties of the boundary-linked expander decomposition sequence in the context of graph cuts, which we will use later on.
First, regardless of the choice whether to add each $u_j$ to $S^{i+1}$, we have the following lemma relating the sets $D^i_j$ to the original set $S$.
\BL\leml{D-lb}
For any decomposition sequence $\{S^i\}$ of $S$,
\[ \pt_GS \s \bigcup_{i=0}^L \bigcup_{j\in[k_i]} \pt_{G^i}D^i_j . \]
\EL
\BP
Observe that
\BG
(\pt_{G^i}S^i) \triangle (\pt_{G^{i+1}}S^{i+1}) \s \bigcup_{j\in[k_i]} \pt_{G^i}D^i_j . \eqnl{diff}
\EG
In particular,
\[ \pt_{G^i}S^i \s \pt_{G^{i+1}}S^{i+1} \cup \bigcup_{j\in[k_i]} \pt_{G^i}D^i_j .\]
Iterating this over all $i$,
\[ \pt_GS \s \bigcup_{i=0}^L  \bigcup_{j\in[k_i]} \pt_{G^i}D^i_j .\]
\EP

We now define a specific decomposition sequence of $S$, by setting up the rule whether or not to include each $u_j$ in $S^i$. For each $U^i_j$, if 
\[ \vol_{G^i[U^i_j]}(S^i\cap U^i_j) + \f\be\phi  w(E_{G^i}(S^i \cap U^i_j, U^i\sm U^i_j)) \ge\vol_{G^i[U^i_j]}(U^i_j\sm S^i) + \f\be\phi  w(E_{G^i}(U^i_j\sm S^i,U^i\sm U^i_j)) ,\] then add $u_j$ to $S^i$; otherwise, do not add $u_j$ to $S^i$. This ensures that
\BG
\vol_{G^i[U^i_j]}(U^i_j\sm D^i_j) + \f\be\phi  w(E_{G^i}(U^i_j\sm D^i_j,U^i\sm U^i_j)) \ge \vol_{G^i[U^i_j]}(D^i_j) + \f\be\phi  w(E_{G^i}(D^i_j, U^i\sm U^i_j))  .\eqnl{Dij}
\EG

Since $G^i[U^i_j]$ is a $\be$-boundary-linked $\phi$-expander, by our construction, we have, for all $i,j$,
\BG
 \f{ w(\pt_{G^i[U^i_j]}D^i_j)}{\vol_{G^i[U^i_j]}(D^i_j) } \ge \phi\eqnl{Exp}
\EG
and
\BG
\f{ w(\pt_{G^i[U^i_j]}D^i_j)}{  w(E_{G^i}(D^i_j, U^i\sm U^i_j))} \ge \be. \eqnl{BL}
\EG

For this specific construction of $\{S^i\}$, called the \emph{canonical} decomposition sequence of $S$, we have the following lemma, which complements \lem{D-lb}.
\BL\leml{D-ub}
Let $\{S^i\}$ be any decomposition sequence of $S$ satisfying \eqn{BL} for all $i,j$. Then,
\[ \sum_{i=0}^L \sum_{j\in[k_i]}  w(\pt_{G^i}D^i_j ) \le\be^{-O(L)} w(\pt_GS) . \]
\EL
\BP
By \eqn{BL},
\[  w(E_{G^i}(D^i_j,U^i\sm U^i_j)) \le\f1\be \cd  w(\pt_{G^i[U^i_j]}D_j^i) .\]
The edges of $\pt_{G^i[U^i_j]}D_j^i$ are inside $\pt_{G^i}S^i$ and are disjoint over distinct $j$, so in total,
\[ \sum_{j\in[k_i]}  w(\pt_{G^i}D^i_j )\le \sum_{j\in[k_i]} \lp1+\f1\be\rp \cd w(\pt_{G^i[U^i_j]}D_j^i ) \le\lp1+\f1\be\rp \cd  w(\pt_{G^i}S^i) .\]
From \eqn{diff}, we also obtain
\[ \pt_{G^{i+1}}S^{i+1} \s \pt_{G^i}S^i \cup \bigcup_{j\in[k_i]} \pt_{G^i}D^i_j .\]
Therefore,
\[  w(\pt_{G^{i+1}}S^{i+1}) \le  w(\pt_{G^i}S^i ) + w\left( \bigcup_{j\in[k_i]} \pt_{G^i}D^i_j \right) \le\lp2+\f1\be\rp \cd  w(\pt_{G^i}S^i) .\]
Iterating this over all $i\in[L]$, we obtain
\[  w(\pt_{G^i}S^i) \le \lp2+\f1\be\rp ^i\cd w(\pt_GS) .\]
Thus,
\[  \sum_{i=0}^L \sum_{j\in[k_i]}  w(\pt_{G^i}D^i_j ) \le \sum_{i=0}^L \f1\be\cd w(\pt_{G^i}S^i) \le \sum_{i=0}^L\f1\be\cd\lp2+\f1\be\rp ^i\cd w(\pt_GS)=\be^{-O(L)} w(\pt_GS) .\]
\EP

\subsection{Unbalanced Case}\secl{unbalanced}

In this section, we generalize the notion of \emph{unbalanced} from \sec{expander} to the general case, and then prove a $(1+\e)$-approximate sparsifier of the unbalanced cuts.

Fix an expander decomposition sequence $\{G^i\}$ of $G$ for the \sec{unbalanced}. For a given set $\emptyset\subsetneq S\subsetneq V$, let $\{S^i\}$ be the canonical decomposition sequence of $S$, and define $D^i_j$ as before, so that they satisfy \eqn{Exp}~and~\eqn{BL} for all $i,j$. We generalize our definition of \emph{unbalanced} from the expander case as follows, for some $\tau=n^{o(1)}$ to be specified later.

\BD
The set $S\s V$ is \emph{$\tau$-unbalanced} if for each level $i$,  $\sum_{j\in[k_i]}\vol_{G^i}(D^i_j)\le\tau\la/\phi$. A cut $\pt S$ is $\tau$-unbalanced if the set $S$ is $\tau$-unbalanced.
\ED
\noindent Note that if $G$ is originally an expander, then in the first expander decomposition of the sequence, we can declare the entire graph as a single expander; in this case, the expander decomposition sequence stops immediately, and the definition of $\tau$-unbalanced becomes equivalent to that from the expander case. We now claim that for an appropriate value of $\tau$, any mincut is $\tau$-unbalanced. 
\BCL\clml{mincut-unbal}
For $\tau\ge\be^{-\Om(L)}$, any mincut $\pt S^*$ of $G$ is $\tau$-unbalanced.
\ECL
\BP
Consider the canonical decomposition sequence of $S$, and define $D^i_j$ as usual.
For each level $i$ and index $j\in[k_i]$,
\BAL
\vol_{G^i}(D^i_j) &= \vol_{G^i[U^i_j]}(D^i_j) + w(E_{G^i}(D^i_j,U^i\sm U^i_j)) 
\\&\stackrel{ \mathclap{ \eqn{Exp}} }\le \f1\phi w(\pt_{G^i[U^i_j]}D^i_j) +w(E_{G^i}(D^i_j,U^i\sm U^i_j)) 
\\&\le  \f1\phi w(\pt_{G^i}D^i_j) .
\EAL
Summing over all $j\in[k_i]$ and applying \lem{D-ub},
\[
 \sum_{j\in[k_i]} \vol_{G^i}(D^i_j) \le  \sum_{j\in[k_i]} \f1\phi w(\pt_{G^i}D^i_j) = \f1\phi\cd \sum_{j\in[k_i]} w(\pt_{G^i}D^i_j ) \stackrel{\textup{Lem.}\ref{lem:D-ub}}\le  \f1\phi\cd\be^{-O(L)} w(\pt_GS^*) \le \f{\tau\la}\phi
 ,\]
 so $S^*$ is $\tau$-unbalanced.
\EP

Let us now introduce some notation exclusive to this section.
For each vertex $v\in U^i$, let $\ol v\s V$ be its ``pullback'' on the original set $V$, defined as all vertices in $V$ that get contracted into $v$ in graph $G^i$ in the expander sequence.
For each set $D^i_j$, let $\ol{D^i_j}\s V$ be the pullback of $D^i_j$, defined as $\ol{D^i_j}=\bigcup_{v\in D^i_j}\ol v$. We can then write
\[ \1_S = \sum_{i,j} \pm\1_{\ol{D^i_j}} = \sum_{i,j}\sum_{v\in D^i_j}\pm\1_{\ol v} ,\]
where the $\pm$ sign depends on whether $D^i_j=U^{i}_j\sm S^i$ or $D^i_j=U^{i}_j\cap S^i$. Then,
\BG
 w(\pt_GS) = \1_S^TL_G\1_S = \sum_{i,j,k,l}\pm\1_{\ol{D^i_j}}^TL_G\1_{\ol{D^k_l}} = \sum_{i,j,k,l} \sum_{u\in D^i_j,v\in D^k_l}\pm\1^T_{\ol u}L_G\1_{\ol v} . \eqnl{ptS}
\EG

\BCL\clml{nonzero}
For an $\tau$-unbalanced set $S$, there are at most $((L+1)\tau/\phi)^2$ nonzero terms in the summation \eqn{ptS}.
\ECL
\BP
Each vertex $v\in D^i_j$ has degree at least $\la$ in $G^i$, since it induces a cut (specifically, its pullback $\ol v\s V$) in the original graph $G$. Therefore,
\[ \tau\la/\phi \ge \sum_{j\in[k_i]}\vol_{G^i}(D^i_j) \ge \sum_{j\in[k_i]}|D^i_j|\cd\la ,\]
so there are at most $\tau/\phi$ many choices for $j$ and $u\in D^i_j$ given a level $i$. There are at most $L+1$ many choices for $i$, giving at most $(L+1)\tau/\phi$ many combinations of $i,j,u$. The same holds for combinations of $k,l,v$, hence the claim.
\EP

The main goal of this section is to prove the following lemma.
\BL\leml{pessimistic}
There exists a constant $C>0$ such that given any weight $W\le \f{C\e^2\phi\la}{\tau\ln(Lm)}$, we can compute, in deterministic $\tO(L^2m)$ time,\footnote{outside of computing the boundary-linked expander decomposition sequence} an unweighted graph $H$ such that for all levels $i,k$ and vertices $u\in U^i,v\in U^k$ satisfying $\deg_{G^i}(u)\le\tau\la/\phi$ and $\deg_{G^k}(v)\le\tau\la/\phi$,
\BG
 \left| \1^T_{\ol u}L_G\1_{\ol v} - W \cd \1^T_{\ol u}L_H\1_{\ol v} \right| \le \e\la . \eqnl{additive}
\EG
\EL
Before we prove \lem{pessimistic}, we show that it implies a sparsifier of $\tau$-unbalanced cuts, which is the lemma we will eventually use to prove \thm{sparsifier-restricted}:
\BL\corl{unbalanced}
There exists a constant $C>0$ such that given any weight $W\le \f{C\e^2\phi\la}{\tau\ln(Lm)}$, we can compute, in deterministic $\tO(L^2m)$ time, an unweighted graph $H$ such that for each \emph{$\tau$-unbalanced} cut $S$,
\[ \big|  w(\pt_GS) - W\cd w(\pt_HS) \big| \le\lp\f{(L+1)\tau}\phi\rp^2\cd\e\la  .\]
\EL
\BP
Let $C>0$ be the same constant as the one in \lem{pessimistic}.
Applying \eqn{ptS} to $\pt_HS$ as well, we have
\[  w(\pt_GS)-W\cd w(\pt_HS) = \sum_{i,j,k,l} \sum_{u\in D^i_j,v\in D^k_l}\pm(\1^T_{\ol u}L_G\1_{\ol v} - W\cd\1^T_{\ol u}L_H\1_{\ol v}) ,\]
so that
\[ \big|  w(\pt_GS)-W\cd w(\pt_HS) \big| \le  \sum_{i,j,k,l} \sum_{u\in D^i_j,v\in D^k_l} \big| \1^T_{\ol u}L_G\1_{\ol v} - W\cd\1^T_{\ol u}L_H\1_{\ol v} \big| .\]
By \clm{nonzero}, there are at most $((L+1)\tau/\phi)^2$ nonzero terms in the summation above. In order to apply \lem{pessimistic} to each such term, we need to show that $\deg_{G^i}(u)\le\tau\la/\phi$ and $\deg_{G^k}(v)\le\tau\la/\phi$. Since $S$ is an $\tau$-unbalanced cut, we have
\[ \deg_{G^i}(u) \le \vol_{G^i}(D^i_j) \le \sum_{j\in[k_i]}\vol_{G^i}(D^i_j) \le \tau\la/\phi  ,\]
and similarly for $\deg_{G^k}(v)$. Therefore, by \lem{pessimistic},
\[ \big|  w(\pt_GS)-W\cd w(\pt_HS) \big| \le \lp\f{(L+1)\tau}\phi\rp^2\cd\e\la ,\]
as desired.
\EP

The rest of \sec{unbalanced} is dedicated to proving \lem{pessimistic}.

Expand out $L_G=\sum_{e\in E}L_e$, where $L_e$ is the Laplacian of the graph consisting of the single edge $e$ of the same weight, so that $\1_{\ol u}^TL_e\1_{\ol v} \in\{-w(e),w(e)\}$ if exactly one endpoint of $e$ is in $\ol u$ and exactly one endpoint of $e$ is in $\ol v$, and $\1_{\ol u}^TL_e\1_{\ol v}=0$ otherwise. 
Let $E_{\ol u,\ol v,+}$ denote the edges $e\in E$ with $\1_{\ol u}^TL_e\1_{\ol v}=w(e)$, and $E_{\ol u,\ol v,-}$ denote those with $\1_{\ol u}^TL_e\1_{\ol v}=-w(e)$.

\subsubsection{Random Sampling Procedure}\secl{random}

Consider the Benzcur-Karger random sampling procedure, which we will de-randomize in this section. Let $\widehat H$ be a subgraph of $G$ with each edge $e\in E$ sampled independently with probability $w(e)/W$, which is at most $1$ by the assumption of \thm{sparsifier-restricted}. Intuitively, the parameter $W\ge\la/f(n)$ is selected so that with probability close to $1$, \eqn{additive} holds over all $i,k,u,v$.

We now introduce our concentration bounds for the random sampling procedure, namely the classical multiplicative Chernoff bound. We state a form that includes bounds on the moment-generating function $\E[e^{tX}]$ obtained in the standard proof. 
\BL[Multiplicative Chernoff bound]\thml{chernoff}
Let $X_1,\lds,X_N$ be independent random variables that take values in $[0,1]$, and 
let $X=\sum_{i=1}^NX_i$ and $\mu=\E[X]=\sum_{i=1}^Np_i$. Fix a parameter $\de$, and define
\BG t^u = \ln(1+\de)\qquad\text{and}\qquad t^l = \ln\lp\f1{1-\de}\rp .\eqnl{t}\EG Then, we have the following upper and lower tail bounds:
\BALN \Pr[X>(1+\de)\mu] &\le e^{-t^u(1+\de)\mu}\E[e^{t^uX}] \le e^{-\de^2\mu/3} , \eqnl{u} \\
\Pr[X<(1-\de)\mu] &\le e^{t^l(1-\de)\mu}\E[e^{-t^lX}] \le e^{-\de^2\mu/3} . \eqnl{l} \EALN
\EL

We now describe our de-randomization by pessimistic estimators. Let $F\s E$ be the set of edges for which a value $X_e\in\{0,1\}$ has already been set, so that $F$ is initially $\emptyset$.  For each $i,k$, vertices $u\in U^i,v\in U^k$, and sign $\circ\in\{+,-\}$ such that $E_{\ol u,\ol v,\circ}\ne\emptyset$, we first define a ``local'' pessimistic estimator $\Phi_{\ol u,\ol v,\circ}(\cd)$, which is a function on the set of pairs $(e,X_e)$ over all $e\in F$. The algorithm computes a $3$-approximation $\widetilde\la\in[\la,3\la]$ to the mincut with the $\tO(m)$-time $(2+\e)$-approximation algorithm of Matula~\cite{matula1993linear}, and sets
\BG
\mu_{\ol u,\ol v,\circ}=\f{ w(E_{\ol u,\ol v,\circ})}{W} \qquad\text{and}\qquad \de_{\ol u,\ol v,\circ}=\f{\e\widetilde\la}{6 w(E_{\ol u,\ol v,\circ})}.\eqnl{mu}
\EG
Following \eqn t, we define
\BG
 t^u_{\ol u,\ol v,\circ}=\ln(1+\de_{\ol u,\ol v,\circ})\qquad\text{and}\qquad t^l_{\ol u,\ol v,\circ}=\ln\lp\f1{1-\de_{\ol u,\ol v,\circ}}\rp, \eqnl{tu}
\EG
 and following the middle expressions (the moment-generating functions) in~\eqn u~and~\eqn l, we define
\BAL
\Phi_{\ol u,\ol v,\circ}(\{(e,X_e):e\in F\}) =  e^{-t^u_{\ol u,\ol v,\circ}(1+\de_{\ol u,\ol v,\circ})\mu_{\ol u,\ol v,\circ}} & \prod_{e\in E_{\ol u,\ol v,\circ}\cap F}e^{t^u_{\ol u,\ol v,\circ}X_e}\prod_{e\in E_{\ol u,\ol v,\circ}\sm F}\E[e^{t^u_{\ol u,\ol v,\circ}X_e}]
\\ +\; e^{t^l_{\ol u,\ol v,\circ}(1-\de_{\ol u,\ol v,\circ})\mu_{\ol u,\ol v,\circ}} & \prod_{e\in E_{\ol u,\ol v,\circ}\cap F}e^{-t^l_{\ol u,\ol v,\circ}X_e}\prod_{e\in E_{\ol u,\ol v,\circ}\sm F}\E[e^{-t^l_{\ol u,\ol v,\circ}X_e}] . 
\EAL
Observe that if we are setting the value of $X_{e'}$ for a new edge $e'\in E_{\ol u,\ol v,\circ}\sm F$, then by linearity of expectation, there is an assignment $X_{e'}\in\{0,1\}$ for which $\Phi_{\ol u,\ol v,\circ}(\cd)$ can only decrease:
\[ \Phi_{\ol u,\ol v,\circ}(\{(e,X_e):e\in F\} \cup (e',X_{e'})) \le \Phi_{\ol u,\ol v,\circ}(\{(e,X_e):e\in F\}) .\]

Since the $X_e$ terms are independent, we have that for any $t\in\R$ and $E'\s E$,
\[ \E\lb e^{t\sum_{e\in E'}X_e}\rb = \Prod_{e\in E'} \E[e^{tX_e}] .\]
By the independence above and the second inequalities in~\eqn u~and~\eqn l, the initial ``local'' pessimistic estimator $\Phi_{\ol u,\ol v,\circ}(\emptyset)$ satisfies
\BAL
\Phi_{\ol u,\ol v,\circ}(\emptyset) \le 2\exp\lp-\f{\de_{\ol u,\ol v,\circ}^2\mu_{\ol u,\ol v,\circ}}3\rp &= 2\exp\lp -\f{(\e\widetilde\la/(6 w(E_{\ol u,\ol v,\circ})))^2\cd w(E_{\ol u,\ol v,\circ})/W}3 \rp \\& = 2\exp\lp-\f{\e^2\widetilde\la^2}{108 w(E_{\ol u,\ol v,\circ}) W}\rp .
\EAL
We would like the above expression to be less than $1$. To upper bound $ w(E_{\ol u,\ol v,\circ})$, note first that every edge $e\in E_{\ol u,\ol v,\circ}$ must, under the contraction from $G$ all the way to $G^i$, map to an edge incident to $u$ in $G^i$, which gives $ w(E_{\ol u,\ol v,\circ})\le\deg_{G^i}(u)$. Moreover, since $\deg_{G^i}(u)\le\tau\la/\phi$ by assumption, we have 
\BG
 w(E_{\ol u,\ol v,\circ})\le\deg_{G^i}(u)\le\tau\la/\phi \eqnl{Euvo}
\EG
 so that
\[ \Phi_{\ol u,\ol v,\circ}(\emptyset) \le  2\exp\lp-\f{\e^2\widetilde\la^2}{108(\tau\la/\phi) W}\rp \le  2\exp\lp-\f{\e^2\la^2}{108(\tau\la/\phi) W}\rp = 2\exp\lp-\f{\e^2\phi\la}{108\tau W}\rp .\]
Assume that
\BG
W \le \f{\e^2\phi\la}{108\tau\ln\lp16(L+1)^2m\rp} ,\eqnl{W}
\EG
which satisfies the bounds in \lem{pessimistic},
so that
\[ \Phi_{\ol u,\ol v,\circ}(\emptyset) \le  2\exp\lp-\f{\e^2\phi\la}{108\tau W}\rp \le \f1{8(L+1)^2m} .\]
Our actual, ``global'' pessimistic estimator $\Phi(\cd)$ is simply the sum of the ``local'' pessimistic estimators:
\[ \Phi(\{(e,X_e):e\in F\}) = \sum_{\substack{i,k,\\u\in U^i,v\in U^k,\\\circ\in\{+,-\}}} \Phi_{\ol u,\ol v,\circ}(\{(e,X_e):e\in F\}) .\]
The initial pessimistic estimator $\Phi(\emptyset)$ satisfies
\[ \Phi(\emptyset) =  \sum_{\substack{i,k,\\u\in U^i,v\in U^k,\\\circ\in\{+,-\}}} \Phi_{\ol u,\ol v,\circ}(\emptyset) \le  \sum_{\substack{i,k,\\u\in U^i,v\in U^k,\\\circ\in\{+,-\}}} \f1{8(L+1)^2m} \stackrel{\text{Clm.}\ref{clm:for-each-edge-2}}\le 4(L+1)^2m\cd\f1{8(L+1)^2m} = \f12 .\]
Again, if we are setting the value of $X_f$ for a new edge $f\in E\sm F$, then by linearity of expectation, there is an assignment $X_f\in\{0,1\}$ for which $\Phi(\cd)$ can only decrease:
\[ \Phi(\{(e,X_e):e\in F\} \cup (f,X_f)) \le \Phi(\{(e,X_e):e\in F\}) .\]
Therefore, if we always select such an assignment $X_e$, then once we have iterated over all $e\in E$, we have
\BG
\Phi(\{(e,X_e):e\in E\} ) \le \Phi(\emptyset)\le\f12 \le 1 . \eqnl{le1}
\EG
This means that for each  $i,k,u\in U^i,v\in U^k$, and sign $\circ\in\{+,-\}$,
\[ \Phi_{\ol u,\ol v,\circ}(\{(e,X_e):e\in E\} ) = e^{-t^u_{\ol u,\ol v,\circ}(1+\de_{\ol u,\ol v,\circ})\mu_{\ol u,\ol v,\circ}} \prod_{e\in E_{\ol u,\ol v,\circ}}e^{t^u_{\ol u,\ol v,\circ}X_e}+\; e^{t^l_{\ol u,\ol v,\circ}(1-\de_{\ol u,\ol v,\circ})\mu_{\ol u,\ol v,\circ}} \prod_{e\in E_{\ol u,\ol v,\circ}}e^{-t^l_{\ol u,\ol v,\circ}X_e} \le1 .\]
In particular, each of the two terms is at most $1$. Recalling from definition \eqn{mu} that $\mu_{\ol u,\ol v,\circ}= w(E_{\ol u,\ol v,\circ})/W$ and $\de_{\ol u,\ol v,\circ}=\e\widetilde\la/(6 w(E_{\ol u,\ol v,\circ}))$, we have
\[ \sum_{e\in E_{\ol u,\ol v,\circ}}X_e \le (1+\de_{\ol u,\ol v,\circ})\mu_{\ol u,\ol v,\circ} = \f{ w(E_{\ol u,\ol v,\circ})}{W}+\f{\e\widetilde\la}{6W}\]
and
\[ \sum_{e\in E_{\ol u,\ol v,\circ}}X_e \ge (1-\de_{\ol u,\ol v,\circ})\mu_{\ol u,\ol v,\circ} = \f{ w(E_{\ol u,\ol v,\circ})}{W}-\f{\e\widetilde\la}{6W} .\]
Therefore,
\[ \left| \1^T_{\ol u}L_G\1_{\ol v} - W \cd \1^T_{\ol u}L_{\widehat H}\1_{\ol v} \right| \le \sum_{\circ\in\{+,-\}} \left| w(E_{\ol u,\ol v,\circ}) - W\cd\sum_{e\in E_{\ol u,\ol v,\circ}}X_e\right| \le \f{\e\widetilde\la}6+\f{\e\widetilde\la}6= \f{\e\widetilde\la}3 \le\e\la , \]
fulfilling \eqn{additive}.

It remains to consider the running time. We first bound the number of $i,k,u,v$ such that either $E_{\ol u,\ol v,+}\ne\emptyset$ or $E_{\ol u,\ol v,-}\ne\emptyset$; the others are irrelevant since $\1_{\ol u}^TL_G\1_{\ol v}=\1_{\ol u}^TL_{\widehat H}\1_{\ol v}=0$. 

\BCL\clml{for-each-edge-0}
For each pair of vertices $x,y$, there are at most $(L+1)^2$ many selections of $i,k$ and $u\in U^i,v\in U^k$ such that $x\in\ol u$ and $y\in\ol v$.
\ECL
\BP
For each level $i$, there is exactly one vertex $u\in U^i$ with $x\in\ol u$, and for each level $k$, there is exactly one vertex $v\in U^k$ with $y\in\ol v$. This makes $(L+1)^2$ many choices of $i,k$ total, and unique choices for $u,v$ given $i,k$.
\EP
\BCL\clml{for-each-edge}
For each edge $e\in E$, there are at most $4(L+1)^2$ many selections of $i,k$ and $u\in U^i,v\in U^k$ and $\circ\in\{+,-\}$ such that $e\in E_{\ol u,\ol v,\circ}$.
\ECL
\BP
If $e\in E_{\ol u,\ol v,\circ}$, then exactly one endpoint of $e$ is in $\ol u$ and exactly one endpoint of $e$ is in $\ol v$. There are four possibilities as to which endpoint is in $\ol u$ and which is in $\ol v$, and for each, \clm{for-each-edge-0} gives at most $(L+1)^2$ choices.
\EP
\BCL\clml{for-each-edge-2}
There are at most $4(L+1)^2m$ many choices of $i,k,u,v,\circ$ such that $E_{\ol u,\ol v,\circ}\ne\emptyset$.
\ECL
\BP
For each such choice, charge it to an arbitrary edge $(x,y)\in E_{\ol u,\ol v,\circ}$. Each edge is charged at most $4(L+1)^2$ times by \clm{for-each-edge}, giving at most $4(L+1)^2m$ total charges.
\EP

By \clm{for-each-edge}, each new edge $e\in E\sm F$ is in at most $4(L+1)^2$ many sets $E_{\ol u,\ol v,\circ}$, and therefore affects at most $4(L+1)^2$ many terms $\Phi_{\ol u,\ol v,\circ}(\{(e,X_e):e\in F\})$. The algorithm only needs to re-evaluate these terms with the new variable $X_e$ set to $0$ and with it set to $1$, and take the one with the smaller new $\Phi(\cd)$. This takes $O(L^2)$ arithmetic operations.

How long do the arithmetic operations take? We compute each exponential in $\Phi(\cd)$ with $c \logn$ bits of precision after the decimal point for some constant $c>0$, which takes $\polylog(n)$ time. Each one introduces an additive error of $1/n^c$, and there are $\poly(n)$ exponential computations overall, for a total of $1/n^c\cd\poly(n)\le1/2$ error for a large enough $c>0$. Factoring in this error, the inequality \eqn{le1} instead becomes
\[ \Phi(\{(e,X_e):e\in E\} ) \le \Phi(\emptyset)+\f12\le\f12+\f12= 1 ,\]
so the rest of the bounds still hold.

This concludes the proof of \lem{pessimistic}.

\subsection{Balanced Case}\secl{lossy}

Similar to the expander case, we treat balanced cuts by ``overlaying" a ``lossy", $n^{o(1)}$-approximate sparsifier of $G$ on top of the graph $\widehat H$ obtained from \cor{unbalanced}. In the expander case, this sparsifier was just another expander, but for general graphs, we need to do more work. At a high level, we compute an expander decomposition sequence, and on each level, we replace each of the expanders with a fixed expander (like in the expander case). Due to the technical proof and lack of novel ideas, we defer the proof to \sec{unweighted}.

\begin{restatable}{theorem}{Lossy}\thml{lossy}
Let $G$ be a weighted multigraph with mincut $\la$ whose edges have weight at most $O(\la)$. For any parameters $\widetilde\la\in[\la,3\la]$ and $\De\ge2^{O(\logn)^{5/6}}$, we can compute, in deterministic $2^{O(\logn)^{5/6}(\log\logn)^{O(1)}}m+O(\De m)$ time, an unweighted multigraph $H$ such that $W\cd H$ is a $\g$-approximate cut sparsifier of $G$, where $\g\le2^{O(\logn)^{5/6}(\log\logn)^{O(1)}}$ and $W=\widetilde\la/\De$. (The graph $H$ does not need to be a subgraph of $G$.) Moreover, the algorithm does not need to know the mincut value $\la$.
\end{restatable}

\subsection{Combining Them Together}

We now combine the unbalanced and balanced cases to prove \thm{sparsifier-restricted}, restated below.

\SpRestricted*

Our high-level procedure is similar to the one from the expander case. For the $\tau$-unbalanced cuts, we use \cor{unbalanced}. For the balanced cuts, we show that their size must be much larger than $\la$, so that even on a $\g$-approximate weighted sparsifier guaranteed by \thm{lossy}, their weight is still much larger than $\la$. We then ``overlay'' the $\g$-approximate weighted sparsifier with a ``light'' enough weight onto the sparsifier of $\tau$-unbalanced cuts. The weight is light enough to barely affect the mincuts, but still large enough to force any balanced cut to increase by at least $\la$ in weight.

\BCL\clml{bal}
If a cut $S$ is balanced, then $ w(\pt _GS)\ge\be^{O(L)}\tau\la$.
\ECL
\BP
Consider the level $i$ for which $\sum_{j\in [k_i]}\vol_{G^i}(D^i_j)>\tau\la/\phi$. For each $j\in[k_i]$, we have
\BAL
 \vol_{G^i}(D^i_j) = \vol_{G^i[U^i_j]}(D^i_j) +  w(E_{G^i}(D^i_j,U^i\sm U^i_j))& \stackrel{\eqn{Exp}}\le\f1\phi w(\pt_{G^i[U^i_j]}D^i_j)+  w(E_{G^i}(D^i_j,U^i\sm U^i_j))
\\&\le\f1\phi\lp w(\pt_{G^i[U^i_j]}D^i_j)+  w(E_{G^i}(D^i_j,U^i\sm U^i_j))\rp
\\&=\f1\phi w(\pt_{G^i}D^i_j) ,
\EAL
so summing over all $j\in[k_i]$,
\[ \sum_{j\in[k_i]}\f1\phi w(\pt_{G^i}D^i_j) \ge  \sum_{j\in[k_i]}\vol_{G^i}(D^i_j) > \f{\tau\la}\phi .\]
By \lem{D-ub}, it follows that
\[  w(\pt_GS) \ge \be^{O(L)}\sum_{j\in[k_i]} w(\pt_G^iD^i_j) \ge \be^{O(L)}\tau\la .\]
\EP

\begin{figure}\centering
\small
    \begin{tabular}{| l | p{12cm} |  }
    \hline
    \ \ \clap{Par.} & Value\\ \hline
$\la$ & Mincut of $G$
\\ \hline
$\widetilde\la$ & $3$-approximation of $\la$
\\ \hline
$\e$ & Given as input
\\ \hline
$r$ & $(\logn)^{1/6}$
\\ \hline
$\be$ & $(\logn)^{-O(r^4)}$ from \thm{exp-decomp}
\\ \hline
$\phi$ & $(\logn)^{-r^5}$ 
\\ \hline
$L$ & $O(\f\logn{r^5})$ 
\\ \hline
$\g$ & $2^{O(\logn)^{5/6}(\log\logn)^{O(1)}}$ from \thm{lossy} 
\\ \hline
$\De$ & $2^{\Th(\logn)^{5/6}}$ from \thm{lossy}
\\ \hline
$\tau$ & $\be^{-cL}\g^2/\e$ for large enough constant $c>0$
\\ \hline
$\e' $ & $\f12(\f\phi{(L+1)\tau})^2\e$
\\ \hline
$\widehat W$ & $\min \{ \f{C\e'\phi\widetilde\la}{\tau\ln(Lm)} , \f{\widetilde\la}\De\} $ where $C>0$ is the constant from \cor{unbalanced}
\\ \hline
$\widetilde W$ & $\f\e{2\g}\cd\f{\widetilde\la}\De$
\\ \hline
    \end{tabular}
\caption{The parameters in the proof of \thm{sparsifier-restricted}.}\label{fig:par}
\end{figure}

We now set some of our parameters; see Figure~\ref{fig:par} for a complete table of the parameters in our proof.
For $r:=(\logn)^{1/6}$, let $\be:=(\logn)^{-O(r^4)}$ and $\phi:=(\logn)^{-r^5}$, so that by \thm{exp-decomp}, the total weight of inter-cluster edges, and therefore the total weight of the next graph in the expander decomposition sequence, shrinks by factor $(\logn)^{O(r^4)}\phi = (\logn)^{-\Om(r^5)}$. Since edge weights are assumed to be polynomially bounded, this shrinking can only happen $O(\f\logn{r^5})$ times, so $L\le O(\f\logn{r^5})$. 

Let $\widetilde\la\in[\la,3\la]$ be a $3$-approximation to the mincut, computable in $\tO(m)$ time~\cite{matula1993linear}, 
Let $\tilde\e := \f12(\f\phi{(L+1)\tau})^2\e$ for parameter $\tau$ that we set later, and let $\widehat H$ be the sparsifier of $\tau$-unbalanced cuts from \cor{unbalanced} for this value of $\tilde\e$ (instead of $\e$) and the following value of $\widehat W\le\f{C\tilde\e^2\phi\la}{\tau\ln(Lm)}$ (taking the place of $W$):
\[ \widehat W:=\min \left\{ \f{C\tilde\e^2\phi\widetilde\la}{3\tau\ln(Lm)} , \f{\widetilde\la}\De \right\} = \min\left\{ \Om\lp\f{\e^2\phi^5\widetilde\la}{\tau^5L^4\ln(Lm)}\rp , \f{\widetilde\la}\De \right\} .\] 
Let $\widetilde H$ be the unweighted graph from \thm{lossy} applied to $\widetilde\la$ and $\De$, so that $\widetilde\la/\De\cd \widetilde H$ is a $\g$-approximate cut sparsifier for $\g:=2^{O(\logn)^{5/6}(\log\logn)^{O(1)}}$. Define $\widetilde W:=\f\e{2\g}\cd\f{\widetilde\la}\De$, and let $H'$ be the ``union" of the graph $\widehat H$ weighted by $\widehat W$ and the graph $\widetilde H$ weighted by $\widetilde W$. More formally, consider a weighted graph $H'$ where each edge $(u,v)$ is weighted by $\widehat W\cd w_{\widehat H}(u,v)+\widetilde W\cd w_{\widetilde H}(u,v)$. 

For an $\tau$-unbalanced cut $\pt S$, the addition of the graph $\widetilde H$ weighted by $\widetilde W$ increases its weight by 
\[ \widetilde W\cd w(\pt_{\widetilde H}S)= \f\e{2\g}\cd\lp\f{\widetilde\la}\De w(\pt_{\widetilde H}S)\rp \le \f\e{2\g}\cd\g w(\pt_GS) = \f\e2 w(\pt_GS) ,\]
so that
\BAL
\left|  w(\pt_GS) - \lp \widehat W\cd w(\pt_{\widehat H}S) +\widetilde W\cd w(\pt_{\widetilde H}S)\rp \right| &\le \big|  w(\pt_GS) - \widehat W\cd w(\pt_{\widehat H}S) \big| +\widetilde W\cd w(\pt_{\widetilde H}S^*) 
\\& \le \lp\f{(L+1)\tau}\phi\rp^2\cd\e'\la + \f\e2 w(\pt_GS)
\\&= \f{\e\la}2+\f\e2 w(\pt_GS)
\\&\le\e w(\pt_GS) .
\EAL
In particular, any $\tau$-unbalanced cut satisfies 
\BG(1-\e)\la\le\widehat W\cd w(\pt_{\widehat H}S) +\widetilde W\cd w(\pt_{\widetilde H}S) \le (1+\e)\la.\eqnl{unbal2}\EG 

Next, we show that all balanced cuts have weight at least $\la$ in the graph $\widetilde H$ weighted by $\widetilde W$. This is where we finally set $\tau:=\be^{-cL}\g^2/\e$ for large enough constant $c>0$. For a balanced cut $S$, 
\[\widetilde W\cd w(\pt_{\widetilde H}S) = \f\e{2\g}\cd\lp\f\la\De w(\pt_{\widetilde H}S)\rp \ge\f\e{2\g}\cd\lp  \f1\g w(\pt_G S)\rp\stackrel{\text{Clm.}\ref{clm:bal}}\ge\f\e{\g^2}\cd\be^{O(L)}\tau\la \ge \la. \]
Moreover, by \clm{mincut-unbal} for this value of $\tau\ge\be^{-O(L)}$, the mincut $\pt S^*$ is $\tau$-unbalanced, and therefore has weight at least $(1-\e)\la$ in $H'$ by~\eqn{unbal2}. 

Therefore, $H'$ preserves the mincut up to factor $\e$ and has mincut at least $(1-\e)\la$. It remains to make all edge weights the same on this sparsifier. Since $\widetilde W=\f\e{2\g}\cd\f{\widetilde\la}\De$ and the only requirement for $\De$ from \thm{lossy} is that $\De\ge2^{O(\logn)^{5/6}}$, we can increase or decrease $\De$ by a constant factor until either $\widetilde W/\widehat W$ or $\widehat W/\widetilde W$ is an integer. Then, we can let $W:=\min\{\widehat W,\widetilde W\}$ and define the unweighted graph $H$ so that $\#_H(u,v)=w_{H'}(u,v)/W$ for all $u,v\in V$. Therefore, our final weight $W$ is 
\BAL W = \min\{\widehat W,\widetilde W\} &= \min\left\{ \Om\lp\f{\e\phi^3\widetilde\la}{\tau^3L^2\ln(Lm)}\rp , \f{\widetilde\la}\De,\f\e{2\g}\cd\f{\widetilde\la}\De \right\}
\\& \ge \e^{7} 2^{-O(\logn)^{5/6}(\log\logn)^{O(1)}} \la  ,\EAL
so we can set $f(n):= 2^{O(\logn)^{5/6}(\log\logn)^{O(1)}} $, as desired.

Finally, we bound the running time. The expander decomposition sequence (\thm{exp-decomp}) takes time $m^{1+O(1/r)}+\tO(m/\phi^2)$, the unbalanced case (\thm{exp-decomp}) takes time $\tO(L^2m)$, and the balanced case takes time $2^{O(\logn)^{5/6}(\log\logn)^{O(1)}}m^{}$. Altogether, the total is $2^{O(\logn)^{5/6}(\log\logn)^{O(1)}}m$, which concludes the proof of \thm{sparsifier-restricted}. 

\subsection{Removing the Maximum Weight Assumption}

Let $f(n)=2^{O(\logn)^{5/6}(\log\logn)^{O(1)}}$ be the function from \thm{sparsifier-restricted}. In this section, we show how to use \thm{sparsifier-restricted}, which assumes that the maximum edge weight in $G$ is at most $\e^7\la/f(n)$, to prove \thm{sparsifier}, which makes no assumption on edge weights.

First, we show that we can assume without loss of generality that the maximum edge weight in $G$ is at most $3\la$. To see why, the algorithm can first compute a $3$-approximation $\widetilde\la\in[\la,3\la]$ to the mincut with the $\tO(m)$-time $(2+\e)$-approximation algorithm of Matula~\cite{matula1993linear}, and for each edge in $G$ with weight more than $\widetilde\la$, reduce its weight to $\widetilde\la$. Let the resulting graph be $\widetilde G$. We now claim the following:
\BCL
Suppose an unweighted graph $H$ and some weight $W$ satisfy the two properties of \thm{sparsifier} for $\widetilde G$. Then, they also satisfy the two properties of \thm{sparsifier} for $G$.
\ECL
\BP
The only cuts that change value between $G$ and $\widetilde G$ are those with an edge of weight more than $\widetilde\la$, which means their value must be greater than $\widetilde\la\ge\la$. In particular, since $G$ and $\widetilde G$ have the same mincuts and the same mincut values, both properties of \thm{sparsifier} also hold when the input graph is $G$.
\EP

For the rest of the proof, we work with $\widetilde G$ instead of $G$. Define $\widetilde W:=\e^7\widetilde\la/(3f(n))$, which satisfies $\widetilde W \le \e^7\la/f(n)$. For each edge $e$ in $\widetilde G$, split it into  $\lc w(e)/\widetilde W\rc$ parallel edges of weight at most $\widetilde W$ each, whose sum of weights equals $w(e)$; let the resulting graph be $\widehat G$. Apply \thm{sparsifier-restricted} on $\widehat G$, which returns an unweighted graph $H$ and weight $W\ge\e^7\la/f(n)$ such that the two properties of \thm{sparsifier} hold for $\widehat G$.  Clearly, the cuts are the same in $\widetilde G$ and $\widehat G$: we have $w(\pt_{\widetilde G}S)=w(\pt_{\widehat G}S)$ for all $S\s V$. Therefore, the two properties also hold for $\widehat G$, as desired.

We now bound the size of $\widehat G$ and the running time. Since $w(e)\le\widetilde\la$, we have $\lc w(e)/\widetilde W\rc \le \lc 3f(n)/\e^7\rc$, so each edge splits into at most $O(f(n)/\e^7)$ edges and the total number of edges is $\widehat m \le O(f(n)/\e^7)\cd m$. Therefore, \thm{sparsifier-restricted} takes time $2^{O(\logn)^{5/6}(\log\logn)^{O(1)}}\widehat m = \e^{-7}2^{O(\logn)^{5/6}(\log\logn)^{O(1)}}m$, concluding the proof of \thm{sparsifier}.




\section*{Acknowledgements}
I am indebted to Sivakanth Gopi, Janardhan Kulkarni, Jakub Tarnawski, and Sam Wong for their supervision and encouragement on this project while I was a research intern at Microsoft Research, as well as providing valuable feedback on the manuscript. I also thank Thatchaphol Saranurak for introducing me to the boundary-linked expander decomposition framework~\cite{goranci2020expander}, as well as Jason Zheng for pointing out multiple technical errors that resulted in an $\e^{-7}$ term instead of $\e^{-4}$ in the statement of \thm{sparsifier}.

\bibliographystyle{alpha}
\bibliography{refs,dp-refs}

\appendix

\section{Boundary-Linked Expander Decomposition}\secl{exp-decomp}

\newcommand{\WBCut}{\textsf{\textup{WeightedBalCutPrune}}}
\newcommand{\bd}{\textbf{\textup{d}}}
\newcommand{\hset}{\mathcal H}
\newcommand{\pset}{\mathcal P}

In this section, we prove \thm{exp-decomp} assuming the subroutine \textsf{WeightedBalCutPrune} from~\cite{LS21}. Our proof is directly modeled off of the proof of Corollary~6.1~of~\cite{CGL19} and the proof of Theorem~4.5~of~\cite{goranci2020expander}, so we claim no novelty in this section.

We will work with weighted multigraphs with \emph{self-loops}, and we re-define the degree $\deg(v)$ to mean $w(\pt(\{v\}))$ plus the total weight of all self-loops at vertex $v$.  All other definitions that depend on $\deg(v)$, such as $\vol(S)$ and $\Phi(G)$, are also affected.

Given a weighted graph $G=(V,E)$, a parameter $r>0$, and a subset $A\s V$, define $G\{A\}^r$ as the graph $G[A]$ with the following self-loops attached: for each edge $e\in E(A,V\sm A)$ with endpoint $v\in A$, add a self-loop at $v$ of weight $r\cd w(e)$. The following observation is immediate by definition:

\BO\label{obs:BL}
For any graph $G=(V,E)$ and subset $A\s V$, Property~(1) of \thm{exp-decomp} holds for $V_i=A$ iff $G[A]^{\be/\phi}$ is a $\phi$-expander.
\EO

We now define the $\WBCut$ problem from~\cite{LS21}\footnote{ Their definition is more general and takes in a \emph{demand} vector $\bd\in\R^V$ on the vertices; we are simply restricting ourselves to $\bd(v)=\deg(v)$ for all $v\in V$, which gives our definition.}
 and their algorithm.

\BD[$\WBCut$ problem, Definition~2.3~of~\cite{LS21}]    
        The input to the $\alpha$-approximate $\WBCut$ problem is a graph $G=(V,E)$, a conductance parameter $0<\phi\leq 1$, and an approximation factor $\alpha$. The goal is to compute a cut $(A,B)$ in $G$, 
        with $w_G(A,B)\leq \alpha \phi\cdot \vol(B)$, such that one of the following holds:
        either 
        \begin{enumerate}
                \item \textup{\textbf{(Cut)}} $\vol(A),\vol(B)\ge \vol(V)/3$; or
                \item \textup{\textbf{(Prune)}} $\vol(A)\geq \vol(V)/2$, and $\Phi(G[A])\ge\phi$.
        \end{enumerate}
\ED

\BT[\WBCut\ algorithm, Theorem~2.4~of~\cite{LS21}]\thml{WBCut}
        There is a deterministic algorithm that, given a graph $G=(V,E)$ with $m$ edges and polynomially bounded edge weights,
        and parameters $0<\psi\leq 1$ and $r\geq 1$, solves the $(\logn)^{O(r^4)}$-approximate $\WBCut$ problem in time $m^{1+O(1/r)}$.
\ET

The \textbf{(Prune)} case requires the additional \emph{trimming} step described in the lemma below.
While \cite{goranci2020expander} prove it for unweighted graphs only, the algorithm translates directly to the weighted case;\footnote{In particular, the core subroutine, called \emph{Unit-Flow} in~\cite{SaranurakW19}, is based on the push-relabel max-flow algorithm, which works on both unweighted and weighted graphs.} see, for example, Theorem~4.2~of~\cite{SaranurakW19}.

\BL[Trimming, Lemmas~4.9~and~4.10 of \cite{goranci2020expander}]\leml{trim}
Given a weighted graph $G=(V,E)$ and subset $A\s V$ such that $G\{A\}$ is an $8\phi$-expander and $w(E_G(A,V\sm A))\le\f\phi{16}\vol_G(A)$, we can compute a ``pruned" set $P\s A$ in deterministic $\tO(m/\phi^2)$ time with the following properties:
 \BE
 \im $\vol_G(P) \le \f4\phi w(E_G(A,V\sm A))$,
 \im $w(E_G(A',V\sm A'))\le2w(E_G(A,V\sm A))$ where $A':=A\sm P$, and
 \im $G\{A'\}^{1/(8\phi)}$ is a $\phi$-expander.
 \EE
\EL

We now prove \thm{exp-decomp} assuming \thm{WBCut}. Our proof is copied almost ad verbatim from the proof of Corollary~6.1~of~\cite{CGL19} on expander decompositions, with the necessary changes to prove the additional boundary-linked property.

        We maintain a collection $\hset$ of vertex-disjoint graphs that we call \emph{clusters}, which are subgraphs of $G$ with some additional self-loops. The set $\hset$ of clusters is partitioned into two subsets, set $\hset^A$ of \emph{active clusters}, and set $\hset^I$ of \emph{inactive clusters}. We ensure that each inactive cluster $H\in \hset^I$ is a $\phi$-expander. We also maintain a set $E'$ of ``deleted'' edges, that are not contained in any cluster in $\hset$. At the beginning of the algorithm, we let $\hset=\hset^A=\{G\}$, $\hset^I=\emptyset$, and $E'=\emptyset$. The algorithm proceeds as long as $\hset^A\neq \emptyset$, and consists of iterations. 
        Let $\al=(\logn)^{O(r^4)}$ be the approximation factor from \thm{WBCut}.

        In every iteration, we apply the algorithm from \Cref{thm:WBCut} to every graph $H\in \hset^A$, with the same parameters $\alpha$, $r$, and $\phi$. Let $U$ be the vertices of $H$. Consider the cut $(A,B)$ in $H$ that the algorithm returns, with 
\BG w(E_H(A,B))\leq \alpha \phi\cdot \vol(U)\leq \frac{\epsilon\cdot \vol(U)}{c\logn} .\eqnl{add-to-E'}\EG We add the edges of $E_H(A,B)$ to set $E'$. 

If $\vol_H(B)\ge \f{\vol(U)}{32\al}$, then we replace $H$ with $H\{A\}^{1/(\al^2\phi\logn)}$ and $H\{B\}^{1/(\al^2\phi\logn)}$ in $\hset$ and in $\hset^A$. Note that the self-loops add a total volume of \BG \f1{\al^2\phi\logn}\cd w(E_H(A,B)) \le \f1{\al^2\phi\logn}\cd\al\phi\,\vol(U) =\f1{\al\logn}\vol(U).\eqnl{selfl}\EG

Otherwise, if $\vol_H(B)<\f{\vol(U)}{32\al}\le\vol(U)/3$, then we must be in the \textbf{(Prune)} case, which means that 
        $\vol_H(A)\geq \vol(U)/2$ and graph $H\{A\}^{1/(8\phi)}$ has conductance at least $\phi$. Since 
\[ w(E_H(A,B))\le\al\phi\cd\vol_H(B) \le \f\phi{32}\vol(U)\le\f\phi{16}\vol(A) ,\]
 we can call \lem{trim} on $A$ to obtain a pruned set $P\s A$ such that 
\[ \vol_H(P) \le \f4\phi w(E_H(A,B)) \le \f18\vol(U) \] and \[ w(E_H(A',U\sm A'))\le2w(E_H(A,B))\le\f\phi8\vol(A) \] for $A':=A\sm P$, and $H\{A'\}^{1/(8\phi)}$ is a $\phi$-expander. Add the edges of $E_H(A',U\sm A')$ to $E'$, remove $H$ from $\m H$ and $\m H^A$, add $H\{A'\}^{1/(8\phi)}$ to $\m H$ and $\m H^I$, and add $H\{B\cup P\}^{1/(8\phi)}$ to $\hset$ and $\hset^A$. Observe that 
\[ \vol_H(B\cup P) = \vol_H(B)+\vol_H(P) \le \f12\vol_H(U) + \f18\vol_H(U)  \le \f58\vol(U) .\]
        
        When the algorithm terminates, $\hset^A=\emptyset$, and so every graph in $\hset$ has conductance at least $\phi$. Notice that in every iteration, the maximum volume of a graph in $\hset^A$ is at most a factor $(1-\f1{32\al})$ of what it was before. Since edge weights are polynomially bounded, the number of iterations is at most $O(\al\log n)$. On each iteration, the total volume of graphs in $\m H^A$ increases by at most factor $1+\f2{\al\logn}$ factor due to the self-loops added in~\eqn{selfl}, so the total volume of all $H\in\m H$ at the end is at most a constant factor of the initial volume $\vol_G(V)$.

The output of the algorithm is the partition of $V$ induced by the vertex sets of $H\in\m H$, so the inter-cluster edges is a subset of $E'$.
 It is easy to verify by~\eqn{add-to-E'} that the total weight of edges added to set $E'$ in every iteration is at most $\al\phi$ times the total volume of graphs in $\m H^A$ at the beginning of that iteration, which is $O(\vol_G(V))$. Over all $O(\al\logn)$ iterations, the total weight of $E'$ is $O(\al\log n)\cd\al\phi\,\vol_G(V) \le (\logn)^{O(r^4)}\phi\vol_G(V)$, fulfilling property~(2) of a boundary-linked expander decomposition.

It remains to show that for each graph $H\in\m H^I$, its vertex set $U$ satisfies the boundary-linked $\phi$-expander property~(1) of \thm{exp-decomp}. For each boundary edge $e\in E_G(U,V\sm U)$, it was created at some iteration where we either added $\f1{\al^2\phi\logn}$ self-loops or $\f1{8\phi}$ self-loops, so $G[U]^{\min\{1/(\al^2\phi\logn),1/(8\phi)\}}$ is a subgraph of $H$. Since $H$ is a $\phi$-expander, so is $G[U]^{\min\{1/(\al^2\phi\logn),1/(8\phi)\}}$, and property~(1) for $\be:=\min\{1/\al^2,1/8\}$ follows by \Cref{obs:BL}.

It remains to analyze the running time of the algorithm.  The running time of a single iteration is bounded by $O( m^{1+O(1/r)})+\tO(m/ \phi^2 )$. Since the total number of iterations is bounded by $O(\logn)$, the total running time is the same, asymptotically.

\section{Lossy Unweighted Sparsifier}\secl{unweighted}

In this section, we prove \thm{lossy}, restated below.

\Lossy*

We will work with weighted multigraphs with \emph{self-loops}, and we re-define the degree $\deg(v)$ to mean $w(\pt(\{v\}))$ plus the total weight of all self-loops at vertex $v$. All other definitions that depend on $\deg(v)$, such as $\vol(S)$ and $\Phi(G)$, are also affected. 

The construction of the sparsifier $H$ is recursive. The original input is graph $G=G^0$, and the output sparsifier will be $H=H^0$. Let the input graph on level $i\ge0$ of the recursion be $G^i$, with $U^i$ as its vertex set. Let $U^i_1,U^i_2,\lds$ be an expander decomposition of $G^i$, and let $G^{i+1}$ be the graph with each set $U^i_j$ contracted to a single vertex $u^{i+1}_j$. If $G^{i+1}$ has more than one vertex, recursively compute a sparsifier on $G^{i+1}$, which still has mincut at least $\la$, and let the sparsifier be $H^{i+1}$. For each edge $(u^{i+1}_j,u^{i+1}_k)$ in $H^{i+1}$, we select a vertex $x\in U^{i}_j$ and $y\in U^{i}_k$ and add edge $(x,y)$ to an initially empty graph $H^i_0$ on $U^i$. We do this in a way that each vertex $v\in U^i_j$ is incident to at most $\left\lc \deg_{H^{i+1}}(u^{i+1}_j) \cd \f{ w(E_{G^i}(v,U^i\sm U^{i}_j))}{\deg_{G^{i+1}}(u^{i+1}_j)} \right\rc$ many edges. Since $\sum_{v\in U^{i}_j} w(E_{G^i}(v,U^i\sm U^{i}_j)) = \deg_{G^{i+1}}(u^{i+1}_j)$, this is always possible by an averaging argument. Next, for each cluster $U^{i}_j$, we compute an $\Om(1)$-expander multigraph $H^{i}_j$ (possibly with self-loops) on the vertices $U^{i}_j$ such that for all $v\in U^{i}_j$, 
\BG
\deg_{G^i}(v)\le W\cd\deg_{H^{i}_j}(v) \le9\deg_{G^i}(v).\eqnl{Hj}
\EG
This can be done by using the lemma below with $d(v)=\deg_{G^i}(v)/W \ge \la/W\ge1$. The running time is at most $O(\sum_{v\in U_i}\deg_{G^i}(v)/W) \le O(\sum_{e\in E}w(e)/W)$, which is $O(m\la/W)=O(\De m)$ since by assumption, all edges in $G$ have weight at most $O(\la)$, and $W=\widetilde\la/\De=\Th(\la/\De)$.
\BL
Given a vertex set $V$ and real numbers $d(v)\ge1:v\in V$, there exists a universal constant $C_0$ such that we can construct, in $O(\sum_{v\in V}d(v))$ time, an $\Om(1)$-expander multigraph $H$ on $V$ (possibly with self-loops) such that for all $v\in V$,
\[ d(v) \le \deg_H(v) \le 9 d(v).\]
\EL
\BP
We use the following theorem of~\cite{CGL19}:
\BT
There is a constant $\al_0>0$ and a deterministic algorithm that, given an integer $n > 1$, in time $O(n)$ constructs a graph $H_n$ with
$|V (H_n)| = n$, such that $H_n$ is an $\al_0$-expander, and every vertex in $H_n$ has degree at most $9$.
\ET
Let $n=\sum_{v\in V}d(v)$, and let $H_n$ be the constructed graph on vertex set $V_n$. Partition $V_n$ arbitrarily into subsets $U_v:v\in V$ such that $|U_v|=d(v)$ for each $v\in V$. Let $H$ be the graph $H_n$ with each set $U_v$ contracted to a single vertex $v$, keeping self-loops, so that $\deg_{H}(v)=\vol_{H_n}(U_v)$. It is not hard to see that expansion does not decrease upon contraction, so $H$ is still an $\Om(1)$-expander. We can bound the degrees $\deg_{H}(v)$ as
\[ d(v)=|U_v|\le\vol_{H_n}(U_v) =\deg_{H}(v)= \vol_{H_n}(U_v) \le 9|U_v| = 9d(v) .\]
\EP

The final sparsifier $H^{i}$ is $H^{i}_0\cup H^{i}_1\cup H^i_2\cup\cds $. This concludes the construction of sparsifier $H^{i}$. (We keep the self-loops, even though they serve no purpose for the sparsifier's guarantees, because we find that including them simplifies the analysis.)
Note that this recursive algorithm implicitly constructs an expander sequence $G^0,G^1,G^2,\lds,G^L$ of $G$ over its recursive calls.

Fix a subset $\emptyset\subsetneq S\subsetneq U^i$, let $S^i$ be the canonical decomposition sequence of $S$, and let $D^i_j$ be constructed as before, so that they satisfy \eqn{Exp}~and~\eqn{BL} for all $i,j$.  

\BCL\leml{deg}
For all $i$ and all $v\in U^i$,
\[ \deg_{G^i}(v) \le W\cd\deg_{H^i}(v) \le 10(L+1)\cd\deg_{G^i}(v) .\]
\ECL
\BP
We prove the stronger statement
\[ \deg_{G^i}(v) \le W\cd\deg_{H^i}(v) \le 10(L+1-i)\cd\deg_{G^i}(v) \]
by induction from $i=L$ down to $0$. For $i=L$, since it is the last level, the entire graph $G^L$ is a single cluster. By construction, $H^L$ consists only of a single constant-expander $H^L_1$ that satisfies $\deg_{G^L}(v)\le W\cd\deg_{H^L_1}(v)\le9\deg_{G^L}(v)$, which completes the base case of the induction.

For $i<L$, by induction, we have $W\cd\deg_{H^{i+1}}(v)\le10(L-i)\cd\deg_{G^{i+1}}(v)$. Fix a cluster $U^i_j$ that  gets contracted to vertex $u^{i+1}_j$ in $G^{i+1}$,   and fix a vertex $v\in U^i_j$. For the graph $H^i_0$, we have 
\BALN
\deg_{H^i_0}(v) \le\left\lc \deg_{H^{i+1}}(u^{i+1}_j) \cd \f{ w(E_{G^i}(v,U^i\sm U^i_j))}{\deg_{G^{i+1}}(u^{i+1}_j)} \right\rc &\le 1+\deg_{H^{i+1}}(u^{i+1}_j) \cd \f{ w(E_{G^i}(v,U^i\sm U^i_j))}{\deg_{G^{i+1}}(u^{i+1}_j)}\nonumber
\\&\le 1+\f{10(L-i)}W w(E_{G^i}(v,U^i\sm U^i_j)) \eqnl{Hi0}
\\&\le1+\f{10(L-i)}W\deg_{G^i}(v). \nonumber
\EALN
For the graph $H^i_j$, by construction \eqn{Hj}, we have
\[ \deg_{G^i}(v)\le W\cd\deg_{H^i_j}(v)\le9\deg_{G^i}(v) .\]
Therefore,
\[ W\cd\deg_{H^i}(v) = W\cd\lp\deg_{H^i_0}(v)+\deg_{H^i_j}(v)\rp \ge W\cd\deg_{H^i_j}(v)\ge\deg_{G^i}(v)\]
and
\[W\cd\deg_{H^i}(v) = W\cd\lp\deg_{H^i_0}(v)+\deg_{H^i_j}(v)\rp \le W+10(L-i)\deg_{G^i}(v) + 9\deg_{G^i}(v) .\]
We can assume that $\De\ge3$, so that $\deg_{G^i}(v) \ge \la \ge\widetilde\la/3=\De W/3\ge W$, and the above is at most
\[ \deg_{G^i}(v)+10(L-i)\deg_{G^i}(v) + 9\deg_{G^i}(v) = 10(L+1-i)\cd\deg_{G^i}(v) ,\]
which completes the induction.
\EP

\BCL[Analogue of \eqn{Exp} for $H^i$]\clml{Exp}
For all $i,j$, 
\[ \f{ w(\pt_{H^i[U^i_j]}D^i_j)}{\vol_{H^i[U^i_j]}(D^i_j)} \ge \Om\lp\f\phi\be\rp .\]
\ECL
\BP
Note that $H^i[U^i_j]$ is exactly $H^i_j$ by construction. We begin by bounding volumes in $G^i$.
\BAL
\vol_{G^i}(D^i_j) &= \vol_{G^i[U^i_j]}(D^i_j) +  w(E_{G^i}(D^i_j,U^i\sm U^i_j))
\\&\le \vol_{G^i[U^i_j]}(D^i_j) + \f\be\phi  w(E_{G^i}(D^i_j,U^i\sm U^i_j))
\\& \stackrel{\eqn{Dij}}\le \vol_{G^i[U^i_j]}(U^i_j\sm D^i_j) + \f\be\phi  w(E_{G^i}(U^i_j\sm D^i_j,U^i\sm U^i_j))
\\&\le\f\be\phi\lp \vol_{G^i[U^i_j]}(U^i_j\sm D^i_j) +  w(E_{G^i}(U^i_j\sm D^i_j,U^i\sm U^i_j) \rp
\\&= \f\be\phi\vol_{G^i}(U^i_j\sm D^i_j).
\EAL
We now translate this volume bound to $H^i_j$. By the construction of $H^i_j$ \eqn{Hj},
\BG
 \vol_{H^i_j}(D^i_j)\le\f{9\vol_{G^i}(D^i_j)}W \le \f{9\be\vol_{G^i}(U^i_j\sm D^i_j)}{W\phi} \le \f{9\be\vol_{H^i_j}(U^i_j\sm D^i_j)}{\phi } .\eqnl{vol}
\EG
Since $H^i_j$ is an $\Om(1)$-expander,
\[  w(\pt_{H^i_j}(D^i_j)) \ge \Om(1) \cd \min\{\vol_{H^i_j}(D^i_j),\vol_{H^i_j}(U^i_j\sm D^i_j)\} \ge \Om(1)\cd\f{\phi}{9\be}\vol_{H^i_j}(D^i_j) ,\]
as desired.
\EP

\BCL[Analogue of \eqn{BL} for $H^i$]\clml{BL}
For all $i<L$ and $j$,
\[ \f{ w(\pt_{H^i[U^i_j]}D^i_j)}{ w(E_{H^i}(D^i_j,U^i\sm U^i_j))} \ge \Om\lp\f1L\rp .\]
\ECL
\BP
Let $u^{i+1}_j\in U^{i+1}$ be the vertex that the cluster $U^i_j$ contracts to in $G^{i+1}$. Again, note that $H^i[U^i_j]$ is exactly $H^i_j$ by construction.

The only edges of $E_{H^i}(D^i_j,U^i\sm U^i_j)$ belong to $H^i_0$, so
\BAL
w(E_{H^i}(D^i_j,U^i\sm U^i_j)) = w(E_{H^i_0}(D^i_j,U^i\sm U^i_j)) &= \sum_{v\in D^i_j}\deg_{H^i_0}(v)
\\&\stackrel{\eqn{Hi0}}\le \sum_{v\in D^i_j} \left(1+\f{10(L-i)}W w(E_{G^i}(v,U^i\sm U^i_j)) \right)
\\&\le \sum_{v\in D^i_j} \left(1+\f{10L}W w(E_{G^i}(v,U^i\sm U^i_j)) \right)
\\&= |D^i_j|+\f{10L}W w(E_{G^i}(D^i_j,U^i\sm U^i_j)).
\EAL
We upper bound the term $ w(E_{G^i}(D^i_j,U^i\sm U^i_j))$ in two ways: 
\BAL
\f\be\phi  w(E_{G^i}(D^i_j,U^i\sm U^i_j))&\le \vol_{G^i[U^i_j]}(D^i_j) + \f\be\phi  w(E_{G^i}(D^i_j,U^i\sm U^i_j))
\\&\le \f\be\phi\lp\vol_{G^i[U^i_j]}(D^i_j)+ w(E_{G^i}(D^i_j,U^i\sm U^i_j))\rp
\\&=\f\be\phi\vol_{G^i}(D^i_j);
\\\f\be\phi  w(E_{G^i}(D^i_j,U^i\sm U^i_j))&\le \vol_{G^i[U^i_j]}(D^i_j) + \f\be\phi  w(E_{G^i}(D^i_j,U^i\sm U^i_j))
\\& \stackrel{\eqn{Dij}}\le \vol_{G^i[U^i_j]}(U^i_j\sm D^i_j) + \f\be\phi  w(E_{G^i}(U^i_j\sm D^i_j,U^i\sm U^i_j))
\\&\le\f\be\phi\lp \vol_{G^i[U^i_j]}(U^i_j\sm D^i_j) +  w(E_{G^i}(U^i_j\sm D^i_j,U^i\sm U^i_j) \rp
\\&= \f\be\phi\vol_{G^i}(U^i_j\sm D^i_j).
\EAL
Therefore,
\BAL
 w(E_{H^i}(D^i_j,U^i\sm U^i_j)) &\le |D^i_j|+\f{10L}W\min\{\vol_{G^i}(D^i_j),\vol_{G^i}(U^i_j\sm D^i_j)\}
\\& \stackrel{\eqn{Hj}}\le |D^i_j|+10L\min\{\vol_{H^i_j}(D^i_j),\vol_{H^i_j}(U^i_j\sm D^i_j)\}\ .
\EAL
We now bound $|D^i_j|$ as follows. By construction \eqn{Hj}, for all $v\in U^i_j$,
\[ \deg_{H^i_j}(v) \ge \f{\deg_{G^i}(v)}W \ge \f\la W\ge\f{\widetilde\la}{3W}=\f\De3 ,\]
which means that
\[ |D^i_j| \le \f{\vol_{H^i_j}(D^i_j)}{\De/3} \stackrel{\eqn{vol}}\le \f{27\be\vol_{H^i_j}(U^i_j\sm D^i_j)}{\De\phi} \le \vol_{H^i_j}(U^i_j\sm D^i_j) \]
as long as we impose the condition
\BG \De\ge\f{27\be}{\phi} \eqnl{De} .\EG
 Therefore,
\[ |D^i_j|\le\min\{\vol_{H^i_j}(D^i_j),\vol_{H^i_j}(U^i_j\sm D^i_j)\} \]
and
\[  w(E_{H^i}(D^i_j,U^i\sm U^i_j))\le(10L+1)\min\{\vol_{H^i_j}(D^i_j),\vol_{H^i_j}(U^i_j\sm D^i_j)\}  .\]

Since $H^i_j$ is an $\Om(1)$-expander,
\[  w(\pt_{H^i_j}D^i_j)\ge\Om(1) \cd \min\{\vol_{H^i_j}(D^i_j),\vol_{H^i_j}(U^i_j\sm D^i_j)\}\ge\Om\lp\f1L\rp\cd  w(E_{H^i}(D^i_j,U^i\sm U^i_j)),\]
as desired.
\EP
\BL\leml{ptDij}
For all $i,j$,
\[ \Om\lp\f\phi\be\rp w(\pt_{G^i}D^i_j)\le W\cd w(\pt_{H^i}D^i_j) \le O\lp\f L\phi\rp w(\pt_{G^i}D^i_j) .\]
\EL
\BP
For the lower bound, we have\BAL
  w(\pt_{H^i}D^i_j) \ge  w(\pt_{H^i[U^i_j]}D^i_j)  \quad&\stackrel{\mathclap{\text{Clm.}\ref{clm:Exp}}}\ge \quad \Om\lp\f\phi\be\rp\vol_{H^i[U^i_j]}(D^i_j) 
\\&\stackrel{\mathclap{\eqn{Hj}}}\ge\Om\lp\f\phi\be\rp\cd\f1W\vol_{G^i}(D^i_j)
\\&\ge \Om\lp\f\phi\be\rp\cd\f1W w(\pt_{G^i}D^i_j)
.\EAL
For the upper bound,
\BAL
 w(\pt_{H^i}D^i_j)&= w(\pt_{H^i[U^i_j]}D^i_j) +  w(E_{H^i}(D^i_j,U^i\sm U^i_j))
\\&\stackrel{\mathclap{\text{Clm.}\ref{clm:BL}}} \le(1+O(L))\cd w(\pt_{H^i[U^i_j]}D^i_j)
\\&\le(1+O(L))\cd\vol_{H^i[U^i_j]}(D^i_j) 
\\&\stackrel{\mathclap{\eqn{Hj}}}\le(1+O(L))\cd \f{9}W \vol_{G^i}(D^i_j)
\\&\le (1+O(L))\cd\f{9}W\lp\vol_{G^i[U^i_j]}(D^i_j)+ w(E_{G^i}(D^i_j,U^i\sm U^i_j))\rp
\\&\le (1+O(L))\cd\f{9}W\lp \f1\phi w(\pt_{G^i[U^i_j]}D^i_j) +  w(E_{G^i}(D^i_j,U^i\sm U^i_j)) \rp
\\&\le (1+O(L))\cd\f{9}W\cd\f1\phi\lp w(\pt_{G^i[U^i_j]}D^i_j) +  w(E_{G^i}(D^i_j,U^i\sm U^i_j)) \rp
\\&= (1+O(L))\cd\f{9}W\cd \f1\phi w(\pt_{G^i}D^i_j) .
\EAL
\EP

\BL
$W\cd H$ is a $\g$-approximate sparsifier with $\g=\max\left\{ O(\be^{-O(L)}L/\phi), O(L^{O(L)}\be/\phi) \right\}$.
\EL
\BP
Since \clm{BL} is an analogue of \eqn{BL} for graph $H$ with the parameter $\be$ replaced by $\Om(1/L)$, we can apply \Cref{lem:D-lb,lem:D-ub} to $H$, obtaining
\BAL
 w(\pt_HS) \le  \sum_{i=0}^L \sum_{j\in[k_i]}  w(\pt_{H^i}D^i_j ) \le L^{O(L)} w(\pt_HS).
\EAL
Combining this with \lem{ptDij},
\BAL  w(\pt_HS) \le  \sum_{i=0}^L \sum_{j\in[k_i]}  w(\pt_{H^i}D^i_j ) &\le(1+O(L))\cd\f{9}W\cd \f1\phi\sum_{i=0}^L \sum_{j\in[k_i]}  w(\pt_{G^i}D^i_j ) \\&\stackrel{\mathclap{\text{Lem.}\ref{lem:D-ub}}}\le \quad(1+O(L))\cd\f{9}W\cd \f1\phi\cd\be^{-O(L)} w(\pt_GS) \EAL
and
\BAL  w(\pt_HS) \ge L^{-O(L)}\sum_{i=0}^L \sum_{j\in[k_i]}  w(\pt_{H^i}D^i_j ) &\ge L^{-O(L)} \cd \sum_{i=0}^L \sum_{j\in[k_i]}   \Om\lp\f\phi\be\rp\f1W w(\pt_{G^i}D^i_j)  \\&\stackrel{\mathclap{\text{Lem.}\ref{lem:D-lb}}}\ge\quad L^{-O(L)}\cd\Om\lp\f\phi\be\rp\f1W w(\pt_GS)  .\EAL
\EP

Finally, we set the parameters $r\ge1,\be,L,\phi$. For $r:=(\logn)^{1/6}$, let $\be:=(\logn)^{-O(r^4)}$ and $\phi:=(\logn)^{-r^5}$, so that by \thm{exp-decomp}, the total weight of inter-cluster edges, and therefore the total weight of the next graph in the expander decomposition sequence, shrinks by factor $(\logn)^{O(r^4)}\phi = (\logn)^{-\Om(r^5)}$. Since edge weights are assumed to be polynomially bounded, this shrinking can only happen $O(\f\logn{r^5})$ times, so $L\le O(\f\logn{r^5})$. Therefore, our approximation factor is
\[ \g=\max\left\{ O(\be^{-O(L)}L/\phi), O(L^{O(L)}\be/\phi) \right\} =O(\be^{-O(L)}L/\phi) = 2^{O(\logn)^{5/6}(\log\logn)^{O(1)}},\]
and the running time, which is dominated by the output size $O(\Delta m)$ and the calls to \thm{exp-decomp} and \lem{trim}, is
\[ O(\Delta m)+m^{1+O(1/r)}+\tO(m/\phi^2) = 2^{O(\logn)^{5/6}(\log\logn)^{O(1)}}m+O(\Delta m) . \]
Finally, the condition $ \De\ge\f{27\be}{\phi}$ from \eqn{De} becomes $\De\ge2^{\Om(\logn)^{5/6}}$, concluding the proof of \thm{lossy}.

\end{document}